\documentclass[10pt,conference]{IEEEtran}

\usepackage{booktabs}
\usepackage{hyperref}
\usepackage{amsmath,amsfonts,amssymb,amsthm, bm}
\usepackage{listings}
\lstdefinelanguage{Coq}{
mathescape=true,
texcl=false,
morekeywords=[1]{Section, Module, End, Require, Import, Export,
  Variable, Variables, Parameter, Parameters, Axiom, Hypothesis,
  Hypotheses, Notation, Local, Tactic, Reserved, Scope, Open, Close,
  Bind, Delimit, Definition, Let, Ltac, Fixpoint, CoFixpoint, Add,
  Morphism, Relation, Implicit, Arguments, Unset, Contextual,
  Strict, Prenex, Implicits, Inductive, CoInductive, Record,
  Structure, Canonical, Coercion, Context, Class, Global, Instance,
  Program, Infix, Theorem, Lemma, Corollary, Proposition, Fact,
  Remark, Example, Proof, Goal, Save, Qed, Defined, Hint, Resolve,
  Rewrite, View, Search, Show, Print, Printing, All, Eval, Check,
  Projections, inside, outside, Def},
morekeywords=[2]{forall, exists, exists2, fun, fix, cofix, struct,
  match, with, end, as, in, return, let, if, is, then, else, for, of,
  nosimpl, when},
morekeywords=[3]{Type, Prop, Set, true, false, option},
morekeywords=[4]{pose, set, move, case, elim, apply, clear, hnf,
  intro, intros, generalize, rename, pattern, after, destruct,
  induction, using, refine, inversion, injection, rewrite, congr,
  unlock, compute, ring, field, fourier, replace, fold, unfold,
  change, cutrewrite, simpl, have, suff, wlog, suffices, without,
  loss, nat_norm, assert, cut, trivial, revert, bool_congr, nat_congr,
  symmetry, transitivity, auto, split, left, right, autorewrite},
morekeywords=[5]{by, done, exact, reflexivity, tauto, romega, omega,
  assumption, solve, contradiction, discriminate},
morekeywords=[6]{do, last, first, try, idtac, repeat},
morecomment=[s]{(*}{*)},
showstringspaces=false,
morestring=[b]",
morestring=[d],
tabsize=3,
extendedchars=false,
sensitive=true,
breaklines=false,
basicstyle=\small,
captionpos=b,
columns=[l]flexible,
identifierstyle={\ttfamily\color{black}},
keywordstyle=[1]{\ttfamily\color{violet}},
keywordstyle=[2]{\ttfamily\color{green}},
keywordstyle=[3]{\ttfamily\color{blue}},
keywordstyle=[4]{\ttfamily\color{blue}},
keywordstyle=[5]{\ttfamily\color{red}},
stringstyle=\ttfamily,
commentstyle={\ttfamily\color{gray}},
literate=
    {\\forall}{{\color{dkgreen}{$\forall\;$}}}1
    {\\exists}{{$\exists\;$}}1
    {<-}{{$\leftarrow\;$}}1
    {=>}{{$\Rightarrow\;$}}1
    {==}{{\code{==}\;}}1
    {==>}{{\code{==>}\;}}1
    {->}{{$\rightarrow\;$}}1
    {<->}{{$\leftrightarrow\;$}}1
    {<==}{{$\leq\;$}}1
    {\#}{{$^\star$}}1
    {\\o}{{$\circ\;$}}1
    {\@}{{$\cdot$}}1
    {\/\\}{{$\wedge\;$}}1
    {\\\/}{{$\vee\;$}}1
    {++}{{\code{++}}}1
    {~}{{\ }}1
    {\@\@}{{$@$}}1
    {\\mapsto}{{$\mapsto\;$}}1
    {\\hline}{{\rule{\linewidth}{0.5pt}}}1
}[keywords,comments,strings]

\lstdefinelanguage{Hol}{
  morekeywords=[1]{val, Define, prove},
  morekeywords=[2]{Type, Prop, Set, true, false, option},
  morecomment=[s]{(*}{*)},
  sensitive=true,
  commentstyle=\color{gray},
  identifierstyle={\ttfamily\color{black}},
  keywordstyle=[1]{\ttfamily\color{violet}},
  keywordstyle=[2]{\ttfamily\color{blue}}
}

\usepackage{todonotes}
\usepackage{bussproofs}                               % proof trees
\usepackage[T1]{fontenc}                              % this is necessary for beramo to work
\usepackage[scaled=0.80]{beramono}% our monospace font
\usepackage{enumitem}
\usepackage[numbers]{natbib}
\usepackage{supertabular}
\usepackage{multirow}

\newcommand{\coq}{Coq}
\newcommand{\hol}{HOL4}

\newcommand{\abs}[1]{\ensuremath{|#1|}}
\newcommand{\meps}[1]{\ensuremath{\varepsilon_{#1}}}

\newcommand{\envRel}{\ensuremath{\sim}}

\DeclareMathOperator\range{\Phi_\mathcal{R}}
\DeclareMathOperator\error{\Phi_\mathcal{E}}
\DeclareMathOperator\type{\Phi_\mathcal{T}}
\DeclareMathOperator\vT{\texttt{\small validTypes}}
\DeclareMathOperator\vFR{\texttt{\small validMachineRanges}}
\DeclareMathOperator\vR{\texttt{\small validRealRange}}
\DeclareMathOperator\vE{\texttt{\small validErrors}}
\newcommand{\fV}{\ensuremath{\mathcal{V}}}
\newcommand{\dV}{\ensuremath{\mathcal{D}}}

\newcommand{\emptyEnv}{\ensuremath{\_ \mapsto \bot}}
\newcommand{\updEnv}[3]{\ensuremath{#1\,[ #2 \mapsto #3]}}

\newcommand{\lemref}[1]{\hyperref[#1]{lemma \ref*{#1}}}
\newcommand{\Lemref}[1]{\hyperref[#1]{Lemma \ref*{#1}}}
\newcommand{\thmref}[1]{\hyperref[#1]{theorem \ref*{#1}}}
\newcommand{\Thmref}[1]{\hyperref[#1]{Theorem \ref*{#1}}}

\newtheorem{theorem}{Theorem}

\begin{document}
\title{A Verified Certificate Checker for\\
Finite-Precision Error Bounds in Coq and HOL4}

\author{\IEEEauthorblockN{Heiko Becker\IEEEauthorrefmark{1},
Nikita Zyuzin\IEEEauthorrefmark{1},
Rapha\"{e}l Monat\footnote{This work was done during an internship at MPI-SWS}\IEEEauthorrefmark{2},
Eva Darulova\IEEEauthorrefmark{1},
Magnus O. Myreen\IEEEauthorrefmark{3} and
Anthony Fox\IEEEauthorrefmark{4}}
\IEEEauthorblockA{\IEEEauthorrefmark{1}MPI-SWS, \IEEEauthorrefmark{2}ENS Lyon, \IEEEauthorrefmark{3}Chalmers University of Technology, \IEEEauthorrefmark{4}University of Cambridge\\
 \IEEEauthorrefmark{1}\{hbecker,zyuzin,eva\}@mpi-sws.org, \IEEEauthorrefmark{2}raphael.monat@ens-lyon.org, \IEEEauthorrefmark{3}myreen@chalmers.se, \IEEEauthorrefmark{4}anthony.fox@arm.com}}


\maketitle

\begin{abstract}
Being able to soundly estimate roundoff errors of finite-precision computations
is important for many applications in embedded systems and scientific computing.
Due to the discrepancy between continuous reals and discrete finite-precision
values,
%Due to the unintuitive nature of finite-precision arithmetic,
automated static analysis tools are highly valuable to estimate roundoff errors.
%for this task.
The results, however, are only as correct as the implementations of the static
analysis tools.
This paper presents a formally verified and modular tool which fully
automatically checks the correctness of finite-precision roundoff error bounds
encoded in a certificate.
We present implementations of certificate generation and checking for both
Coq and HOL4 and evaluate it on a number of examples from the literature.
The experiments use both in-logic evaluation of Coq and HOL4, and execution
of extracted code outside of the logics: we benchmark Coq extracted
unverified OCaml code and a CakeML-generated verified binary.

%%% Local Variables:
%%% mode: latex
%%% TeX-master: "paper"
%%% End:

\end{abstract}
%
% !TEX root = paper.tex
\section{Introduction}
\label{sec:intro}

Numerical programs, common in scientific computing or embedded systems, are
often implemented in finite-precision arithmetic.
This approximation of real
numbers inevitably introduces roundoff errors, potentially making the computed
results unacceptably inaccurate. The discrepancy between discrete
finite-precision arithmetic and continuous real arithmetic
%unintuitive nature of finite-precision
%arithmetic as well as the discrepancy between its finite nature and
%continuous real arithmetic
make accurate and sound error estimation
challenging. Automated tool support is thus highly valuable.

This fact was already recognized previously and resulted in a number of static
analysis techniques and
tools~\cite{Goubault2011,Solovyev2015,Darulova2014,Daumas2010} for computing
sound worst-case absolute error bounds on numerical errors. The results of such
static analysis tools are, however, only as correct as the tools' implementation.

%\heiko{Can someone proof-read the paragraph below? I tried to address some of the comments by reviewer 2}
Some of these tools provide independently checkable formal proofs,
however we found that none of the current certificate producing tools,
FPTaylor~\cite{Solovyev2015}, PRECiSa~\cite{moscatostatic} and
Gappa~\cite{Daumas2010} go far enough.
% FPTaylor produces a proof certificate in HOL-Light, but due to the limited
% functionality of HOL-Light, checking each certificate is highly inefficient.
%FPTaylor produces a proof certificate in HOL-Light, but its analysis is specific
%to floating-point arithmetic and does not support other finite precisions.
FPTaylor produces a proof certificate in HOL-Light, relying on an in-logic decision procedure~\cite{solovyev2013formal}.
Its analysis is specific to floating-point arithmetic and does not support other finite precisions.
PRECiSa and Gappa generate a proof certificate by instantiating library theorems, explicitly encoding verification steps.
%cannot be extended to other finite precisions, such as fixed-point arithmetic.
%PRECiSa and Gappa generate a proof script, which explicitly encodes all
%verification steps.
Any tool that explicitly encodes verification steps, or is to be used
interactively~\cite{de2006assisted,Ramananandro2016} requires expert knowledge
in IEEE754 floating-point semantics~\cite{ieee75408} or formal verification; in contrast our goal is to
make our tool usable by non-experts.
% Thus, these tools as well as tools to be used
% interactively~\cite{de2006assisted,Ramananandro2016}
% require expert knowledge in formal verification; in contrast our goal is to make
% our tool usable by non-experts.
Finally, in-logic verification of certificates %and especially proof scripts
can often become unreasonably slow.

This paper describes a new fully automated tool, called \emph{FloVer}, which
checks %\emph{proof certificates}%
proof certificates of finite-precision roundoff error bounds
generated by static analysis tools.
Certificates checked by FloVer encode only the minimal static analysis result,
and thus using FloVer does not require formal verification expertise.
Separately from FloVer, we implement fully automated certificate generation in
the static analysis tool Daisy~\cite{darulova2018daisy}, demonstrating
our envisioned tool-chain.

FloVer supports straight-line arithmetic kernels, floating-point as well as
fixed-point arithmetic, mixed-precision evaluation (including floating-point type inference),
and local variable declarations.
For floating-point expressions, FloVer proves correctness of each
analyzed expression with respect to the concrete bit-level IEEE754
floating-point semantics~\cite{ieee75408}.
Our tool is formally verified in both \coq{} and \hol{}.
A succesful run of FloVer shows that the encoded roundoff error is a valid
upper bound and that the analyzed function can be run without any errors (e.g.
division-by-zero).

In order to handle both floating-point and fixed-point arithmetic, FloVer supports a forward
dataflow static analysis.
FloVer is furthermore built modularly to allow reusability and easy
extensions, and supports dataflow analysis with both interval and affine arithmetic
abstract domains.

We have implemented and verified FloVer in two theorem provers
% partly for the extra assurance, partly to compare the two provers, but mostly
to be able to connect to projects in both provers and thereby make FloVer widely
applicable. In \coq{}, we hope to link to the CompCert
compiler~\cite{Leroy-Compcert-CACM} and CertiCoq~\cite{certicoq}; and in
\hol{} we already link to CakeML~\cite{icfp16}.

%\heiko{We can put more motivation here if we have space, i.e. CompCert C vs. CakeML}

The connection to CakeML allows us to provide efficient certificate checking:
using the CakeML toolchain~\cite{icfp16,myreen2012proof} we produce a verified
binary of our certificate checker. At the time of writing, CertiCoq was not
capable of extracting our checker functions, thus we extract an unverified
binary from \coq{} and compare its perfomance with the verified CakeML binary.

Our evaluation on standard benchmarks from embedded systems and scientific
computing shows that roundoff errors verified by FloVer are competitive with
the state of the art, and extracted certificate checking times are significantly
faster than in-logic verification.

\subsection*{Contributions}
\begin{itemize}
\item We explain our modular, fully automated and self-contained approach to
  certification of absolute finite-precision roundoff error bounds (Section~\ref{sec:checker_details} and \ref{sec:soundness}).
\item We implement and prove FloVer correct in both \coq{} and \hol{}.
  %This allows reuse of our results when combining FloVer with other developments
  %in these theorem provers.
  The sources are available at \url{https://gitlab.mpi-sws.org/AVA/FloVer}.
\item We are the first to provide an efficient and verified way of checking
  finite-precision error certificates by extracting a verified binary version of
  FloVer from \hol{} (\autoref{sec:extraction}).
  %The verified binary
  %is produced by in-logic compilation using the CakeML compiler toolchain.%\vspace{0.5em}
\item We experimentally evaluate (in~\autoref{sec:experiments}) implementations
  of FloVer on examples from the literature. The results are competitive and
  show that our approach to certificate checking is feasible. During our
  experiments, we found a subtle bug in the Daisy static analyzer.
\end{itemize}
%%% Local Variables:
%%% mode: latex
%%% TeX-master: "paper"
%%% End:

%
% !TEX root = paper.tex
\section{Overview}

In this section, we give a high-level overview of our certificate generation and
checking approach. The next section provides the necessary background on finite
precision arithmetic and static dataflow analysis for roundoff errors.
\autoref{sec:checker_details} describes the technical details of FloVer.

A certificate (in \coq{} or \hol{}) checked by FloVer encodes the \emph{result}
of a forward dataflow static analysis of roundoff errors, but not the analysis
or correctness proofs themselves. For each analyzed arithmetic expression
(consisting of $+,-,*,/$, \texttt{\small FMA}, and local variables), the certificate contains:
\begin{itemize}
  \item the expression $f$, as an abstract syntax tree (AST)
  \item a precondition $P$, specifying the domain (interval) of all input variables
  \item a (possibly mixed-precision) type assignment $\Gamma$ for all input
    variables and optionally let-bound variables,
  \item the analysis result which consists of a range $\range$ and an error
        bound $\error$ for each intermediate subexpression
    %(i.e. AST node)
\end{itemize}

FloVer then checks the analysis result recursively, by verifying for each AST
node that the error bound is a sound upper bound on the worst-case absolute
roundoff error:
\begin{align}
\label{eqn:absError}
\max_{x \in [a, b]}\;\;\lvert f(x) - \tilde{f}(\tilde{x}) \rvert
\end{align}
where $f$ and $x$ are the real-valued expression and variable, respectively,
and $\tilde{f}$ and $\tilde{x}$ their finite-precision counterparts.
The interval $[a,b]$ is the domain of $x$ given by precondition $P$.
Ranges for input variables as well as the analysis result are necessary as
(absolute) finite-precision roundoff errors depend on the magnitude of the
computed values. In the absence of input ranges, roundoff errors are unbounded
in general.

\begin{figure}
  \begin{center}
    \includegraphics[scale=.2]{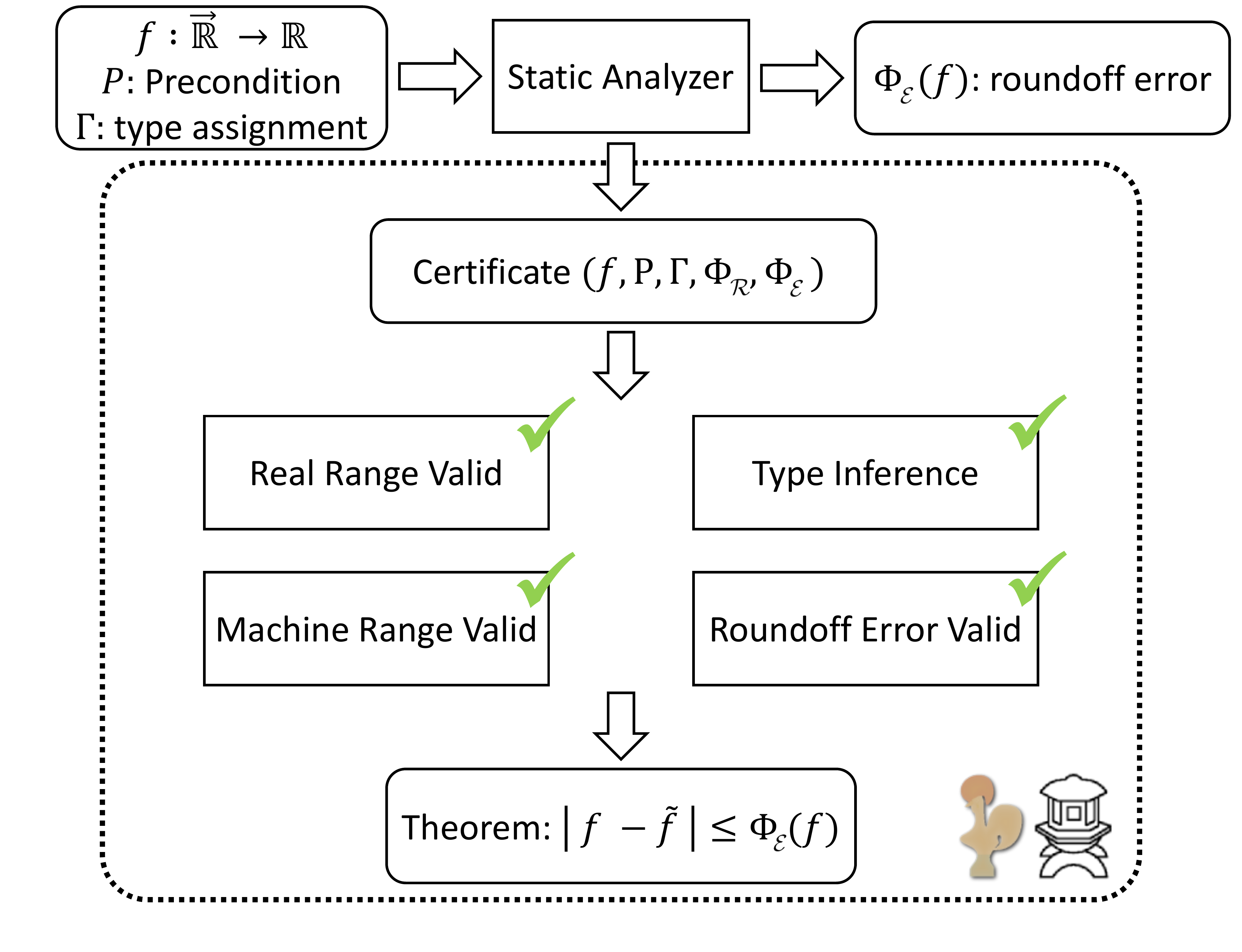}
    \caption{Overview of the FloVer framework}
    \label{fig:overview_structure}
  \end{center}
\end{figure}

FloVer splits the certification into several subtasks and runs separate
validator functions (see also~\autoref{fig:overview_structure}):
\begin{itemize}
  \item $\vR$ validates the range result $\range$,
  \item $\vT$ infers and checks types (given in $\Gamma$) of all
    subexpressions
  \item $\vE$ validates the error results $\error$,
  \item $\vFR$ validates that no overflow and NaN's (not-a-number special
    values) occur.
\end{itemize}

We have implemented the validators in both \coq{} and \hol{} and proven an
overall soundness theorem: when all validators return successfully, then the
computed error bounds (for each subexpression) are soundly overapproximating the
finite-precision roundoff errors. %wrt. to IEEE754 semantics.
%This architecture is also depicted in~\autoref{fig:overview_structure}.

To verify a certificate, one can run the validator functions in \coq{} or \hol{}
directly. However, while both provers natively support evaluation of functions,
this is not particularly efficient. To speed up the certificate checkers, we
have used the CakeML in-logic compilation toolchain~\cite{myreen2012proof}, to
extract a \emph{verified} binary from our \hol{} checker definitions. Since the
CakeML compiler is fully verified, the binary enjoys the very same correctness
guarantees as the certificate checkers implemented in \hol{}. Similarly, we have
used the extraction mechanism~\cite{letouzey2002new} in \coq{} to extract an,
albeit unverified, binary. The binary implementations of the checkers run
natively and are thus significantly more efficient, as our experiments
in~\autoref{sec:experiments} demonstrate.

% then the background on floats etc.

%%% Local Variables:
%%% mode: latex
%%% TeX-master: "paper"
%%% End:

%
% !TEX root = paper.tex
\section{Background}

\subsection{Finite-Precision Arithmetic}\label{subsec:fp_arith}

FloVer uses a general abstraction for finite-precision arithmetic relating it to
operations on real numbers:
\begin{align}
  \label{eqn:fp_comp}
  x \circ_{\mathit{fp}} y = (x \circ y) + \operatorname{error} (x \circ y,
  \mathit{fp}) %\;\text{, }\quad \lvert \delta \rvert \le \meps{fp}
\end{align}
where $\circ \in \{+, -, *, /\}$ and $\circ_{\mathit{fp}}$ denotes the corresponding
finite-precision operation at type $\mathit{fp}$.
Function $\operatorname{error} (e, \mathit{fp})$ computes the error from
representing the real value $e$ in the finite-precision type $\mathit{fp}$.
An input $x$ may not be representable in finite-precision arithmetic, and thus
FloVer considers an initial error on the input:
$\lvert x - \tilde{x} \rvert \leq \operatorname{error} (x,\mathit{fp})$.

For floating-point arithmetic, we assume IEEE754~\cite{ieee75408} semantics
with rounding-to-nearest rounding mode and the standard
abstraction of arithmetic operations:
\begin{align}
\label{eqn:floats}
  \operatorname{error} (e, fp) = e * \delta \quad \lvert \delta \rvert \le \meps{fp}
\end{align}
Constant \meps{fp} is the so-called machine epsilon for precision $fp$ ($fp =$
16, 32 or 64 bits) and represents the maximum relative error for a single
arithmetic operation.
In addition to binary operations, FloVer also supports unary negation, which
does not incur a roundoff error, and fused-multiply-add instructions, where
$\operatorname{FMA} (x,y,z)_{\mathit{fp}} = (x * y + z) + \operatorname{error} (x * y + z, \mathit{fp})$.

\autoref{eqn:floats} holds under IEEE754 floating-point semantics only for
normal floating-point values, and thus FloVer reports ranges containing only
subnormals, infinity or not a number (NaN) special values as errors. We discuss
the proof of correctness wrt. to IEEE754 semantics in~\autoref{subsec:IEEE_soundness}.

Fixed-point arithmetic is an alternative to floating-points which does not
require dedicated hardware and is thus a common choice in embedded systems. No
standard exists, but we follow the common representation~\cite{Anta2010} of fixed-point values
as bit vectors with an integer and a fractional part, separated by an implicit
radix point, which has to be precomputed at compile-time. We assume truncation
as the rounding mode for arithmetic operations. The absolute roundoff error at
each operation is determined by the fixed-point format, i.e. the (implicit)
number of fractional bits available, which in turn can be computed from the
range of possible values at that operation. Since this information must be
computed by any static analysis on fixed-point programs, we encode fractional
bits as part of our fixed-point type and rely on the certificate containing a
full (unverified) map $\Gamma$ from expressions to types for
%rely on the certificate containing a map $F$ from expressions to fractional bits for
fixed-point kernels.

\subsection{Static Dataflow Roundoff Error Analysis}

FloVer's range and error validators perform dataflow roundoff error
analysis and for this follow the same approach for computing absolute error
bounds as Rosa~\cite{Darulova2014}, Fluctuat~\cite{Goubault2011},
Gappa~\cite{Daumas2010} and Daisy~\cite{darulova2018daisy}.

The magnitude of absolute finite-precision roundoff errors depends on the
magnitude of values of all intermediate subexpressions (this can be seen e.g.
from~\autoref{eqn:floats}). Thus, in order to accurately bound roundoff
errors, the analysis first needs to be able to bound the ranges of all
(intermediate) expressions.

% One may think that just evaluating the program in interval arithmetic and
% interpreting the width of the resulting interval as the error bound would be
% sufficient. While this is certainly a sound approach, it computes too
% pessimistic error bounds in general. This is especially true if we consider
% relatively large ranges on inputs; we have no way of distinguishing which part
% of the interval width is due to the input interval or due to accumulated
% roundoff errors. Hence we need to compute ranges and errors separately.

At a conceptual level, dataflow analysis computes roundoff error bounds in two steps:
\begin{description}
  \item[range analysis]
    computes sound range bounds for all intermediate expressions,
  \item[error analysis]
    propagates errors from subexpressions and computes the new worst-case roundoffs
    using the previously computed ranges.
\end{description}
Both steps are performed recursively on the AST of the arithmetic expressions.
A side effect of this separation is that it provides us with a modular approach:
we can choose different range arithmetics with different accuracy-efficiency
tradeoffs for ranges and errors.
Common choices for range arithmetics are interval arithmetic
(IA)~\cite{Moore1966} and affine arithmetic(AA)~\cite{Figueiredo2004}.

% Interval arithmetic (IA)~\cite{Moore1966} is a common efficient choice for range
% estimation, which computes a bounding interval for each basic operation as
% \begin{equation*}
% x \circ^{\#} y = [ \min(x \circ y), \max( x \circ y)], \quad\circ \in \lbrace +, -, *, /  \rbrace
% \end{equation*}
% Affine arithmetic (AA)~\cite{Figueiredo2004} is more expensive to compute, but tracks
% linear correlations between variables and thus can be more accurate.
% As an initial proof-of-concept we implement both interval arithmetic
% and affine arithmetic based checking in $\vR$.

%%% Local Variables:
%%% mode: latex
%%% TeX-master: "paper"
%%% End:

%
% !TEX root = paper.tex
\section{Certification of Error Analysis Results}\label{sec:checker_details}
% High-level certification idea
\begin{figure*}
  \begin{minipage}{.46\textwidth}
  \begin{lstlisting}[frame=single,mathescape,language=Coq,basicstyle=\footnotesize\ttfamily]
  Definition f:cmd Q := <AST f>.
  (* Type assignment for free variables *)
  Definition Gamma: expr Q $\rightarrow$ option mType := <$\Gamma$>.
  (* range constraints on free variables of f *)
  Definition Precondition: nat $\rightarrow$ (Q $*$ Q) := <P>.
  (* map from sub-expressions to ranges and errors *)
  Definition AbsEnv: expr Q $\rightarrow$ option ((Q $*$ Q) $*$ Q) :=
    <$\range,\error$>.

  Theorem CertificateCheckingSucceds =
   CertificateChecker f Gamma Precond AbsEnv = true.
  Proof.
    vm_compute; auto.
  Qed.
  \end{lstlisting}
  \end{minipage}
  \begin{minipage}{.02\textwidth}
  \hspace{0.02\textwidth}
  \end{minipage}
  \begin{minipage}{.46\textwidth}
  \begin{lstlisting}[basicstyle=\footnotesize\ttfamily,frame=single,mathescape,language=Hol]
  val f_def = Define `f: real cmd = <AST f>`;
  (* Type assignment for free variables *)
  val Gamma_def = Define
    `Gamma: real expr $\rightarrow$ mType option = <$\Gamma$>`;
  (* range constraints on free variables of f *)
  val Precondition_def = Define `
    P: num $\rightarrow$ (real $*$ real) = <P>`;
  (* map from sub-expressions to ranges and errors *)
  val AbsEnv_def = Define `
    AbsEnv: real expr $\rightarrow$ ((real $*$ real) $*$ real) option =
      <$\range,\error$>`;
  val CertificateCheckingSucceeds = prove (
   ``CertificateChecker f Gamma Precond AbsEnv``,
   daisy_eval_tac);
  \end{lstlisting}
  \end{minipage}
  \caption{Certificate structure with corresponding types in \coq{} (left) and \hol{} (right)}
  \label{fig:pseudocert}
\end{figure*}
%%% Local Variables:
%%% mode: latex
%%% TeX-master: "paper"
%%% End:

Next, we focus on the technical details of our certificate checking.
The certificates in \coq{}, \hol{} and for the extracted binaries are
structurally the same and only differ in syntax.
\autoref{fig:pseudocert} shows a sample structure of a certificate in \coq{} and
\hol{}, including the types of encoded results.
$\Gamma$ represents a type assignment to all free variables in the analyzed
function.
Expressions (of type \lstinline{expr}) are parametric in the type of constants.
$\range$ and $\error$ map each AST node of the analyzed function to an interval
and a positive (absolute) error bound represented by a single fraction, respectively.
We discuss the differing types of $\range$ and $\error$ in
\autoref{subsec:formal_details}.

The validator functions, which check the certificate, also have the same
structure in both \coq{} and \hol{} and we describe them here independently of the
particular prover.

% abstract domains for checking
\subsection{Checking Range Analysis Results}
The range validator is implemented in the function \\$\vR (e, P, \range)$ which
takes as input an expression $e$, the precondition $P$, which captures the
constraints on the input variables, and the real-valued ranges which are to be
checked in $\range$.
$\vR$ verifies by structural recursion on the AST that for each
subexpression $e'$ of $e$, $\range(e')$ returns a sound enclosure of the
true range, which is computed inside the theorem prover with interval or affine
arithmetic.
That is, we check the ranges in $\range$ by effectively recomputing them inside
the prover.

Since FloVer supports let-bindings in the input program to reuse evaluation
results, both at runtime as well as in the certificate validator, we extend
$\vR$ to handle let-bound variables without recomputing results.
%handle let-bindings without recomputing results.

% mixed-precision support
\subsection{Mixed-precision Support}
Mixed-precision evaluation allows different arithmetic operations to be
executed in different precisions. This often allows to speed up computations as
evaluation in lower precisions is usually faster. Instead of requiring e.g.
uniform 64-bit precision, each subexpression in FloVer can be evaluated in
16, 32 or 64 bit floating-point precisions (each with the corresponding machine
epsilon $\meps{p}$).
FloVer supports the same semantics as C and Scala: for two operands with
different precisions, the lower one is implicitly cast to the higher precision,
but an explicit cast is required when decreasing precision (e.g. when assigning
a 64 bit value to a 32 bit variable).

The typing environment $\Gamma$ assigns a machine precision to every free
variable of the analyzed expression. % to the precision at which it is to be evaluated.
We further require any constant in the AST, as well as casts to be annotated
with its (resulting) precision.

For floating-point precisions, FloVer infers the remaining types automatically,
i.e. the user only has to provide this necessary minimal information, and in
particular does not need to annotate all intermediate operations.

We can reuse the existing infrastructure to support fixed-point arithmetic.
A fixed-point type in FloVer is then represented as a pair of word
length $w$ and number of fractional bits $f$. For fixed-point precisions, FloVer
avoids recomputing the fractional bits and thus relies on the information being
encoded in $\Gamma$.
% As FloVer internally infers the types, we want to avoid recomputing types.
% To this end, we have implemented a function computing a type-map $\type$
% which maps each subexpression of the analyzed function to a
% precision.
% As we do not require a fully type-annotated AST of the analyzed function,
% the precisions are inferred internally by FloVer.
% In order to avoid recomputing types, we have implemented a function computing a
% type-map $\type$, which maps each subexpression of the analyzed function to a
% precision.

When checking a certificate, FloVer computes a full type-map $\type$ from the
(partial) type map $\Gamma$ to avoid recomputing results. To this end we
implement the function $\vT(\Gamma, e)$. The function returns $\type$
if and only if all types encoded in $\Gamma$ are valid types for their
respective subexpressions.
We reuse $\type$ in both  the error validator and the machine range validator.
% When checking a certificate, FloVer first computes the complete type-map $\type$,
% to avoid recomputing results.
% When computing a type-map for floating-point precisions, $F$ is ignored. For
% fixed-point computations $F$ is used to determine the fractional bits of binary
% operations and FMA's, as these operations might change the fractional bits part
% of the result type.
% The type-map is then given together with the type environment $\Gamma$, the
% optional map to fractional bits $F$, and the analyzed function to the
% \emph{type validator} $\vT$ in the certificate checking pipeline.
% $\vT$ returns true if and only if the types encoded in the type-map are correct.
% If no valid type-map could be computed from the given type annotations, FloVer fails.
% If $\vT$ succeeds, $\type$ is reused in both the error validator and the machine range validator.

% generalized error model
\subsection{Checking Error Analysis Results}\label{subsec:check_exp}
The error validator $\vE (e, \type, \range, \error)$ takes as input the
expression $e$, a type assignment to subexpressions $\type$, the range analysis
result $\range{}$ and the error analysis result $\error{}$, which is to be
checked.% and the set of defined variables $\dV$.
That is, $\vE$ assumes that the ranges and types have been verified
independently.
As for the range validator,  we extend $\vE$ to reuse results of let-bound
variables. %uses the set $\dV$ to track previously checked let-bound variables.
The validator function checks by structural recursion on the AST of $e$ that for
each subexpression $e'$ of $e$, $\error(e')$ is a sound upper bound on the
absolute roundoff error.

For constants and variables, the error bounds are straight-forwardly derived
using \autoref{eqn:fp_comp} and the range analysis result.
For arithmetic operations, the error check is more involved.
% and we explaine it here
% for the case of addition: $e = e_1 + e_2$ in detail.
% In our notation, the absolute error is given by~\autoref{eqn:absError} as
% \[
% \abs{e - \tilde{e}} = \abs{(e_1 + e_2) - (\tilde{e}_1 +_{\mathit{fl}} \tilde{e}_2)}.
% \]
% Then, using the abstraction of floating-point arithmetic from \autoref{eqn:floats} we obtain
% \[
%   \tilde{e}_1 +_{\mathit{fl}} \tilde{e}_2 = \tilde{e}_1 + \tilde{e}_2 + (\tilde{e}_1 + \tilde{e}_2 ) * \delta,
%   \quad{}\abs{\delta} \leq \meps{p}.
% \]
Using \autoref{eqn:fp_comp}, \autoref{eqn:absError} and the triangle inequality,
we obtain for an addition:
%Combining the two equations and applying the triangle inequality, we obtain
\begin{align}\label{eqn:ex_roundoff}
  \begin{split}
    \abs{(e_1 + e_2) - &(\tilde{e}_1 +_{\mathit{fp}} \tilde{e}_2)} \leq\\
    &\abs{e_1 - \tilde{e}_1} + \abs{e_2 - \tilde{e}_2} +
    \operatorname{error} ((\tilde{e}_1 + \tilde{e}_2), fp)
  \end{split}
\end{align}
$\abs{e_1 - \tilde{e}_1}$ and $\abs{e_2 - \tilde{e}_2}$ are the roundoff errors of the operands,
which are propagated simply by addition.
$\operatorname{error} ((\tilde{e}_1 + \tilde{e}_2), fp)$ is the new roundoff
error commited by the addition at precision $fp$.
The new roundoff error depends on the magnitude of the operands and thus on the
ranges of $\tilde{e}_1$ and $\tilde{e}_2$.

%\heiko{I have cut the explanation on how to upper bound with IA as this is pretty straight forward}
% To upper bound the right hand side of~\autoref{eqn:ex_roundoff}, we use the
% following two properties of intervals.
% %
% \begin{lemma}\label{lem:maxabs_ub}
%   \[
%     a \in [a_{lo},a_{hi}] \Rightarrow a \leq \maxAbs{a_{lo}}{a_{hi}}
%   \]
% \end{lemma}
% \Lemref{lem:maxabs_ub} states that for each interval, the maximum absolute value
% of the outer points is an upper bound on any element in the interval.

% \begin{lemma}\label{lem:perturb_iv}
%   \[
%     \abs{a - \tilde{a}} \leq e \wedge a \in [a_{lo},a_{hi}] \Rightarrow \tilde{a} \in [a_{lo} - e, a_{hi} + e]
%   \]
% \end{lemma}
% The second property (\Lemref{lem:perturb_iv}) states that if we know the
% error on $a$ and $a$'s interval enclosure, we can compute an
% interval for the corresponding floating-point value by widening the bounds of
% the interval by the error.
% We have proven both lemmas in \coq{} and \hol{}.

The computation of an upper bound to~\autoref{eqn:ex_roundoff} then uses the
range analysis result from $\range$, the already verified error bounds on the
subexpressions $e_1$ and $e_2$ in $\error$, and basic properties of range arithmetic.

Similar bounds can be derived for the other arithmetic operations.
However, for multiplication and division, the propagation of errors is more
involved.
For $e_1 * e_2$ we obtain %by \autoref{eqn:floats} and the triangle inequality
$
  \abs{(e_1 * e_2) - (\tilde{e_1} *_{\mathit{fp}}\tilde{e_2})} \leq \abs {e_1 * e_2 - \tilde{e_1} * \tilde{e_2}} + \operatorname{error}(\tilde{e_1} * \tilde{e_2}, fp)
$
and similarly for division:
\begin{align*}
  \abs{(e_1 / e_2) - (\tilde{e_1} /_{\mathit{fp}} \tilde{e_2})} &{\leq}\\
  \abs {e_1 * (1/e_2) - \tilde{e_1} &* (1/\tilde{e_2})} +
  \operatorname{error}(\tilde{e_1} * 1/\tilde{e_2}, fp)
\end{align*}

FloVer checks whether a division by zero may occur during the execution of the analyzed
function under the real-valued as well as the finite-precision semantics.
If it detects that a division by zero can occur in any of the executions, certificate checking
fails.

\subsection{Supported Range Arithmetics}
FloVer currently supports interval arithmetic (IA)~\cite{Moore1966} in both
provers and affine arithmetic (AA)~\cite{Figueiredo2004} in Coq to check
real-valued ranges.
The support for AA in the error validator in Coq as well as the HOL4
development in general is currently work in progress.
Arithmetic operations in IA are efficiently computed as: %s a bounding interval for each basic operation as
%\begin{equation*}
$
x \circ^{\#} y = [ \min(x \circ y), \max( x \circ y)], \quad\circ \in \lbrace +, -, *, /  \rbrace.
$
%\end{equation*}
IA cannot track correlations between variables (e.g. it cannot show that $e_1 - e_1 \in [0,0]$).
Affine arithmetic is a simple relational analysis which tracks linear correlations and thus computes
ranges for linear operations exactly (like the $e_1 - e_1$); for nonlinear operations
it nonetheless has to compute an over-approximation.
% We further discuss the performance differences of interval and affine arithmetic
% in \autoref{sec:experiments}.
%%% Local Variables:
%%% mode: latex
%%% TeX-master: "paper"
%%% End:

%
% !TEX root = paper.tex
\section{Soundness}\label{sec:soundness}
%
%As soundness theorem for FloVer,
We have proven in both \coq{} and \hol{} that
it suffices to run the validator functions on a certificate to show
a) that the static analysis result is correct, and b) that the analyzed function
will always evaluate to a finite value.
% We call this property the soundness of the validator functions.
The overall soundness proof relates a succeeding run of the validators
$\vT$, $\vR$, $\vFR$ and $\vE$ to the semantics of the analyzed function.

We have formalized the semantics of functions according to \autoref{eqn:fp_comp}.
The rule for binary addition, for instance, is %given in \autoref{fig:bin_eval}.
\begin{prooftree}
  \AxiomC{$m_+ = m_1 \sqcup m_2$} %\abs{\delta} \leq \meps{m_+} \qquad$}
  \noLine
  \UnaryInfC{$\type (e_1) = m_1 \quad \type (e_2) = m_2 \quad
    \type (e_1 + e_2) = m_+$}
  \noLine
  \UnaryInfC{$(e_1,E,\type) \Downarrow (v_1, m_1) \qquad
    (e_2,E,\type) \Downarrow (v_2,m_2)$}
  \UnaryInfC{$(e_1 + e_2, E,\type) \Downarrow
    ((v_1 + v_2) + \operatorname{error} (v_1 + v_2, \,m_+))$}
\end{prooftree}
$E$ is the environment tracking values of bound variables, and $\Gamma$ tracks
precisions of variables.
$(e_1,E,\type) \Downarrow (v_1, m_1)$ means that expression $e_1$ big-step
evaluates for the variable environment $E$ and the type assignment $\type$ to
value $v_1$ in precision $m_1$.
%As we support mixed-precision, evaluation returns both the result of evaluating
%the expression, as well as the precision of the returned value.
$m_1 \sqcup m_2$ is an upper bound operator on precisions, returning the
most precise of $m_1$ and $m_2$.

% As the type system supports fixed-point types, fixed-point computations are
% supported by the semantics.
% To support fixed-point evaluations we add the fractional bits map $F$ to the
% semantics ($F$ is ignored for floating-point executions).
% We further add the side condition that if the result types $m_1$ and
% $m_2$ are fixed-point types, then $F(e_1 + e_2)$ must return some number of
% fractional bits $f$ where the join of $m_1$ and $m_2$ has exactly $f$ fractional
% bits.

Real-valued executions map every variable, constant and cast operation to infinite
(real-valued) precision, which we denote by $m = \infty$.
The rules for subtraction, multiplication, division, casts, and FMA's are
defined analogously.
Unary negation does not introduce a new roundoff error and keeps the precision
of the operand.

Analogously to expressions, we will use $E$ to refer to the idealized real-valued
environment and $\tilde{E}$ for the finite-precision environment.
The overall soundness theorem is then% given in \autoref{fig:flover_sound}.

\begin{theorem}
  \label{thm:flover_sound}
  Let $f$ be a real-valued function, $E$ a real-valued environment,
  $\tilde{E}$ its finite-precision counterpart, $P$ a
  precondition constraining the free variables of $f$, $\Gamma$ a map from all
  free variables of $f$ to a precision, $\range$ a
  range analysis result, $\type$ a type-map and $\error$ an error analysis result.
  Then
  \begin{gather*}
    E  \sim_{(\error,\fV,\dV, \type)} \tilde{E}\ \wedge\\
    \vT(\Gamma, f) = \type\ \wedge \vR (f, P, \range)\ \wedge\\
    \vFR(f, \type, \range, \error)\ \wedge \\
    \vE (f, \type, \range, \error)\ \Longrightarrow{}\\
    \exists~v\,\tilde{v}_1\,m_1.\;
    (f,E,\type) \Downarrow (v,\infty)\ \wedge
    (\tilde{f}, \tilde{E}, \type) \Downarrow (\tilde{v}_1,m_1)\ \wedge\ \\
    (\forall \tilde{v}_2\,m_2.\;
    (\tilde{f}, \tilde{E}, \type) \Downarrow (v_2,m_2) \Rightarrow
    \abs{v - \tilde{v}_2} \leq \error(f))
  \end{gather*}
\end{theorem}

\noindent The assumption $E  \sim_{(\error,\fV,\dV, \type)} \tilde{E}$ states
that the real-valued environment $E$ and the finite-precision environment
$\tilde{E}$ agree up to a fixed $\delta$ on the values of the variables in the
sets $\fV$ and $\dV$. We give the full explanation when explaining soundness
of the error validator.
To prove the theorem, we have split the proof into
separate soundness proofs for each validator function. Each theorem is shown by
structural induction on $e$.

%------------ Type validator
\paragraph{Type Validator}
Giving the full type map $\type$ is tedious to do for a user. FloVer thus
requires only annotations for casts, constants and (let-bound) variables, and
infers the remaining types ($\type$) fully automatically for floating-point
expressions. For fixed-point types only, we require $\Gamma$ to be a complete
map since we rely on the fractional bits to be inferred externally.
%an annotation with the fractional bits in $F$ for
%fixed-point types only.

Soundness of the type inference $\vT$ %, checking the type assignment $\type$,
means that when $\type (e) = m_t$ and evaluation of $e$ gives value $v$ and
precision $m_v$, then $m_t = m_v$.
% We found it easier to show the soundness of a separate type validator than of
%the type inference function.
%Also,
Thus, we need not recompute type information once the type map has been computed
and reuse it in the other validators.

% ----------- Range validator
\paragraph{Real Range Validator}
For $\vR$, the soundness theorem proves that if $E$ binds variables in $e$ to
values that are within the range given by the precondition $P$, then
 $e$ evaluates for environment $E$ to $v$ under a real-valued semantics
 and $v$ is contained in $\range(e)$.
 % To prove soundness of $\vR$ we had to prove
% monotonicity of interval arithmetic (\Lemref{lemma:iv_arith_mon}) for each
% supported binary operator and negation.
% \begin{lemma}
%   \label{lemma:iv_arith_mon}
%   Let $I$ and $J$ be intervals and $a,b$ real values. Then
%   \[
%     a \in I \wedge b \in J \Rightarrow (a \circ b) \in I \circ^\# J
%   \]
% \end{lemma}
% \heiko{Cut the monotonicity lemma. If we have space, we should consider adding it back with an explanation that this is the crucial thing to prove for ANY new range arithmetic}

% ----------- FPRange Validator
\paragraph{Machine Range Validator}
We prove that whenever $\vFR$ succeeds on expression $e$, valid type-map $\type$
and valid error map $\error$, then any evaluation of $e$ results
in a finite, representable value for the type of $e$ in $\type$.

For floating-point precisions this means that $v$ is a finite value according
to IEEE754  (i.e. either 0, subnormal or normal).
For fixed-point precisions with word size $w$ and $f$ fractional bits, this
means that $v$ is within the range of representable values
($\abs{v} \leq \frac{2^{w-1}-1}{2^f}$) and no overflow occurs (i.e. the fractional
bits were correctly inferred).

FloVer uses \autoref{eqn:floats} to compute an error for floating-point
precisions which is only valid in the presence of IEEE754
\emph{normal} numbers or $0$. We note that the roundoff error of the
\emph{biggest representable subnormal number} is smaller than the roundoff
error of \emph{normal numbers} in general.
We add this condition as a check to function $\vFR$ by checking that the
floating-point range contains at least one normal number.

% ------------  Error validator
\paragraph{Error Validator}
If $\vE (e, \type, \range, \error)$ succeeds, and $e$ evaluates to $v$, then we want to
show that $\tilde{e}$ evaluates to $\tilde{v}$, and that $\abs{v - \tilde{v}}
\leq \error (e)$. The challenge in this proof lies in the fact that we reason
about two different executions of similar expressions, $e$ and $\tilde{e}$.

Given a free variable $x$ in the analyzed expression $e$,
the value $E (x)$ may not be representable as a finite-precision value.
Thus the values for the related variables $x$ and $\tilde{x}$ will not in general
agree.
This is the case for every free variable occurring in $e$. Additionally, the
roundoff error of any variable depends on its precision.
As a consequence we introduce an inductive approximation
relation $\envRel_{(\fV,\type)}$ between values provided by $E$ and $\tilde{E}$
for variables in $\fV{}$ so that we can prove the error bound.
Given $E \envRel_{(\fV{}, \type)} \tilde{E}$, both environments are defined for
every variable $v \in \fV{}$. In addition, the difference between $E (v)$ and
$\tilde{E} (\tilde{v})$ at precision $p$ is upper bounded by
$\operatorname{error}(v, p)$, where $p$ is $\type (v)$.
In the proofs we instantiate $\fV$ by the free variables of the
analyzed expression.
Two empty environments are trivially related under the empty set
($(\emptyEnv) \envRel_{(\emptyset,\,\type)} (\emptyEnv)$) and for free
variables we have:
% is the set of free
% variables of $e$ and $\Gamma$ a type assignment to them, the environments $E$ and
% $\tilde{E}$ are related on the free variables of $e$ and that each free variable
% is in the domain of both environments.
% \todo[inline]{Eva: there seems to be some duplication in the explanation of $\envRel{}$,
% I would put this explanation before the rules?}
% The rules for constructing empty environments and for free variables
% are:
\begin{prooftree}
  \AxiomC{$E \envRel_{(\fV,\type)} \tilde{E} \qquad{} x \not\in \fV$}
  \noLine
  \UnaryInfC{$\type (x) = m \qquad \abs{\,v - \tilde{v}\,}\:\leq\:\operatorname{error}(v,m)$}
  \LeftLabel{FreeVar}
  \UnaryInfC{($\updEnv{E}{x}{v}) \envRel_{(\{x\} \cup \fV, \type)}
    (\updEnv{\tilde{E}}{\tilde{x}}{\tilde{v}})$}
\end{prooftree}
To prove soundness for let-bindings, we will extend the relation with a rule for
defined variables later.

$\error$ maps expressions to rationals, representing absolute error bounds.
FloVer computes error bounds from intervals from $\range$ and the error bounds on subterms.
The propagation errors for multiplication and division
%Note that the propagation error $\abs{e_1 * e_2 - \tilde{e_1} * \tilde{e_2}}$
depend on both the real-valued and the float-valued ranges. Therefore the
soundness proof requires solving 16 and 32 sub-cases for multiplication and
division, respectively.

% ---------------- Let bindings
\paragraph{Let-Bindings}
To extend the soundness proofs to let-bindings, we have to check that the
analyzed function $f$ is in SSA form (since $\type$, $\range$ and $\error$ are maps, variables cannot be redefined). For this we use the formalization of SSA defined in the LVC
framework~\cite{SchneiderSH15}. %since it nicely states SSA form using sets of
%variables.
% \eva{CUT? We have implemented a simple function that checks whether the encoded function is
% in SSA form and fails otherwise. We have proven in both \coq{} and \hol{}, that
% if the check can be run successfully on function $f$, then $f$ is in SSA form.}
%
Furthermore, we adapt the approximation relation $\envRel$ to include let-bound
variables:
% \eva{CUT? Take \dLet{m}{x}{e}{g}. If $\error$ has been validated for $e$, and the check
% $\error(x) = \error (e)$ succeeds, the roundoff error $\abs{x - \tilde{x}}$ is
% upper bound by $\error (x)$.
% %Hence during the execution of $f$, the floating point value to which $\tilde{x}$
% %is bound is at most off by $\error (x)$.
% We extend $\envRel{}$ by a rule that allows us to record this information inside
% the relation:}
\begin{prooftree}
  \AxiomC{$E \sim_{(\error,\fV,\dV, \type)} \tilde{E} \qquad x \not\in \fV \cup \dV$}
  \noLine
  \UnaryInfC{$\type(x) = m  \qquad \abs{\,v - \tilde{v}\,}\:\leq\:\error\,(x)$}
  \LeftLabel{DefinedVar}
    \UnaryInfC{$(\updEnv{E}{x}{v}) \sim_{(\error,\fV,\{x\} \cup \dV, \type)}
      (\updEnv{\tilde{E}}{\tilde{x}}{\tilde{v}})$}
  \end{prooftree}
\noindent Set $\dV{}$, tracks variables added to both environments using let-bindings
and $\error$ is the error analysis result.
The sets \dV{} and \fV{} are used to distinguish whether a variable $x$ is free
or let-bound.

% --------------- Tying together
\paragraph{Using Flover}
We obtain the overall soundness of FloVer (\autoref{thm:flover_sound}) as the
conjunction of the results of the functions $\vT$, $\vR$, $\vFR$ and $\vE$.
\autoref{thm:flover_sound} holds only if checking of the certificate succeeds.
If the static analysis result in a certificate is incorrect, e.g.
a computed range or roundoff error is incorrect,
FloVer fails checking the certificate.
% \eva{CUT? If FloVer fails checking the certificate in-logic, the final theorem in the
% certificate is not provable. If an extracted binary cannot check a certificate,
% it returns a non-zero return code and prints ``False''.}
Our tool can be used by any other roundoff error analysis tool that computes
real-valued ranges, roundoff error bounds and knows about variable types.
Using FloVer is then as easy as implementing a pretty-printer for this information.

FloVer performs sound dataflow analysis, which necessarily computes an
overapproximation of the true roundoff errors.
It is thus possible that FloVer cannot verify a certificate even though the
error bounds are indeed correct.
Different range arithmetics, which influence the accuracy of FloVer's analysis,
commit different overapproximations.
Thus we use our implementations of IA and AA in Coq in a portfolio approach and
run both when checking range analysis results.
%One common alternative to interval
%arithmetic is affine arithmetic, however, we would like to note that it is not
%in general stronger for nonlinear arithmetic.

% it may always be the case that FloVer cannot
% verify a certificate, because the range bound or the roundoff error cannot be
% validated although they may be correct when using a stronger analysis than interval arithmetic (i.e. affine-arithmetic~\cite{Figueiredo2004})

% Checking of the certificate soundly fails if a type cannot be inferred
% correctly, a real-valued range bound is unsound, a roundoff error bound is wrong
% or a roundoff error cannot be validated inside FloVer (although it may be
% correct in other approaches).
%FloVer will soundly fail checking the certificate and report this.

% ---------------- Floating-Point range validator
\subsection{Connecting FloVer to IEEE754}\label{subsec:IEEE_soundness}

% We want to connect FloVer's formalization to IEEE754 by relating evaluations
We connect our formalization to formalizations of IEEE754 floating-point arithmetic in
HOL4~\cite{IEEE_FP_HOL4} and the Flocq library in Coq~\cite{Boldo2011}
by proving that if checking the certificate succeeds, we can evaluate
the analyzed function using IEEE754 semantics and the roundoff error bound
is valid for this execution.

\begin{theorem}
  \label{thm:IEEE_conn}
  Let $\tilde{f}$ be a function on 64-bit floating-points and $f$ its
  real-valued counterpart, $E$ a real-valued environment, $\tilde{E}$ its 64-bit
  floating-point counterpart, $P$ a precondition constraining the free variables
  of $\tilde{f}$, $\Gamma$ a map from all free variables of $\tilde{f}$ to
  64-bit precision, $\range$ a range analysis result, and $\error$ an error
  analysis result,
  Then
  \begin{gather*}
    E  \sim_{(\error,\fV,\dV, \Gamma)} \tilde{E}\;\wedge\\
    \texttt{CertificateChecker}\,(f, P, \Gamma, \range, \error)\;\wedge\\
    \texttt{IEEEevalAvoidsSubnormals}\,\tilde{f}\;\Longrightarrow\\
    \exists\,v\,\tilde{v}.\;(f,E) \Downarrow v\ \wedge\ (\tilde{f},\tilde{E})
    \Downarrow_{\text{IEEE}} \tilde{v}\ \wedge\
    \abs{v - \tilde{v}} \leq \error(f)
  \end{gather*}
\end{theorem}

% We use the result of $\vFR$ to show necessary precondition for applying lemmas
% from the library
\noindent The proof of \autoref{thm:IEEE_conn} is an extension of FloVer's
soundness theorem (\autoref{thm:flover_sound}). To show that the roundoff error
bounds are valid for the IEEE754 operations, we use the soundness theorem of
$\vFR$ to establish that all values obtained form an evaluation are finite.

% As HOL4 (currently) does not support casts/subnormals, we add the side
% condition that evaluation is done in double precision and no evaluation
% might return a subnormal
The formalization in HOL4 (currently) does support neither cast operations
nor reasoning about roundoff errors for subnormal values.
Until these are supported, we assume $\Gamma$ to map every variable to 64-bit
double precision and disallow subnormal values to occur during evaluation.
To this end, we define the function
$\texttt{\small IEEEevalAvoidsSubnormals}(e, E)$, as a temporary workaround.
The function returns true only if every subexpression of $e$ evaluates to a
normal value or $0$.

% --------------- Bug found
\subsection{Division Bug Found}
We use Daisy~\cite{darulova2018daisy} to generate certificates for our evaluation.
During this, we found a subtle bug in the tool's static analysis of
the division operator.
The error bounds are only sound in the absence of division-by-zero
errors, but only the real-valued range of the denominator was checked for
whether it contains zero. It is possible, however, that the real-valued range
does \emph{not} contain zero, while the corresponding floating-point range does,
essentially due to large enough roundoff errors.

\subsection{Formalization Details}\label{subsec:formal_details}

% The input of FloVer is encoded in the certificate as the AST
% of the analyzed function using the following grammar:
% \begin{align*}
%   x &\in \mathbb{N} \quad{}c \in \mathbb{V} \quad{}m \in \{16,32,64\}\\
%   e_1,e_2\:&::=\:x\:|\:c\:|\:- e_1\:|\:(m) e_1 \:|\:e_1 \circ e_2 \quad{}\circ \in \{+,-,*,/\}\\
%   f\:&::=\:\dLet{x}{e}{f}\:|\:e
% \end{align*}
% where the let-bindings encode local variable declarations.
% % Since the input
% % to \daisy{} is a valid Scala program, let-bindings appear only at the top level.
% Variables $x$ are encoded as natural numbers. The definition is parametric in the
% type of constants $\mathbb{V}$.

% \todo[inline]{reviewer 3: which formalization of floating-point arithmetic
% do you use in Coq/HOL4?\\ Heiko: Sufficient explanation?\\
% Eva: yes, but can you include references/URLs? also I would not say necessary, but
% say that we use them (we could have done them ourselves too...)}
Executions inside FloVer are represented in both \coq{} and \hol{} as big-step
relations using \autoref{eqn:fp_comp}. These formalizations do not depend on
external libraries. Only the connection to IEEE754 semantics uses external libraries.
%External libraries are only used, when proving the
%connection to IEEE754 semantics.
%To this end, we use the formalization by
%\citeauthor{IEEE_FP_HOL4}~\cite{IEEE_FP_HOL4} in \hol{} and the Flocq library by
%\citeauthor{Boldo2011}~\cite{Boldo2011} in \coq{}.

Roundoff errors and our theorems relate real-valued executions to finite-precision
ones and we thus need a way to represent the numbers and also compute on them.
However, the latter is problematic for infinite-precision reals.
% \daisy{} uses internally rationals ($\mathbb{Q}$) implemented with big integers
% for the numerator and denumerator and
We use rationals to represent the values in the certificates.
To relate these values to the real-valued ($\mathbb{R}$) executions in the
theorem statement, we use the fact that rationals are a subset of the
\verb!real! type in \hol{}, and in \coq{} we use the translation
\verb!Q2R!:$\mathbb{Q} \rightarrow \mathbb{R}$ and exploit that our AST is
parametric in the constant type by instantiating it with $\mathbb{Q}$ for
computations and $\mathbb{R}$ for theorems.
%%% Local Variables:
%%% mode: latex
%%% TeX-master: "paper"
%%% End:

%
% !TEX root = paper.tex
\section{Extracting a Verified Binary with CakeML}\label{sec:extraction}
Running the range and error checker functions in \coq{} and \hol{} directly
is quite inefficient (see our experiments in~\autoref{sec:experiments}).
We have thus extracted a verified binary from our \hol{} checker function
definitions, and an unverified binary for \coq{}.
We are aware of the work on certified extraction from \coq{} in the
CertiCoq~\cite{certicoq} project, but at the time of writing, the tool could
not handle our checker definitions.

We have implemented in \hol{} and \coq{} an unverified lexer and parser for
the encoding of the certificates, which are included in the extracted binaries 
in both \coq{} and \hol{}.

\paragraph{Extracting from \hol}
For extracting a binary from \hol{}, we use the CakeML proof-producing synthesis
tool~\cite{myreen2012proof} which translates ML-like \hol{} functions
into deeply embedded CakeML programs that exhibit the same behaviour.
In \hol{} we use the \lstinline!real! type to store the rational bounds in $\range$ and $\error$.
For each of the arithmetic operations over the \lstinline!real! type
that we used in the \hol{} development, we define a translation into
a representation of the arbitrary-precision rationals in CakeML.
% Each HOL \lstinline!real! is represented as a pair of arbitrary precision integers $n$ and $d$ that
% implement the rational $\frac{n}{d}$, where $d$ is positive.
%CakeML's integers are arbitrary precision integers.

CakeML and \hol{} have different notions of equality.
Since we perform equality tests in the certificate checkers, we had to prove that
our newly defined representation of real numbers respects CakeML's semantics
for structural equality. For this purpose, we had to require and
prove that our representation of rationals maintains a gcd of one between nominator
and denominator.
% $gcd (n,d) = 1$.

When translating a \hol{} function into CakeML code, the CakeML toolchain generates
preconditions that exclude runtime exceptions, e.g. divisions by zero.
We have shown that all generated preconditions are always satisfied, hence the
specification theorem for the generated ML code does not have any unproved preconditions left.

Having compiled the CakeML libraries beforehand, we can compile the checking
functions into a verified binary in around 90 minutes on the same machine as we
used for the experiments in~\autoref{sec:experiments}.
Checking the certificate with the binary is then extremely fast, since no theorem
prover logic is loaded.

\paragraph{Extracting from \coq{}}
\coq{} natively supports unverified extraction into OCaml code~\cite{letouzey2002new}.
We used the existing libraries for translating \coq{} numbers into OCaml's
\verb!Big_int! type from the base library.
The extracted code is compiled using the OCaml native-code compiler (``ocamlopt'') in our experiments.

%%% Local Variables:
%%% mode: latex
%%% TeX-master: "paper"
%%% End:

%
%!TEX root = paper.tex
% \item We have extracted a binary using cakeML extractor
%   With the cakeML extractor we extracted a SML definition of our certificate checker in HOL4 and then compiled it with CakeML to a binary.
%   Running checkers in logic is inefficient, therefore we extracted the binary
% \item Experiments: Non-linear benchmarks create peeks
\section{Evaluation}\label{sec:experiments}
{\small
\begin{table}
  \begin{center}
    \renewcommand*{\arraystretch}{1.2}
    \begin{tabular}{lrrr}
      \toprule
  Benchmark & \multicolumn{2}{c}{FloVer} & FPTaylor\\
   & \multicolumn{1}{c}{interval} & \multicolumn{1}{c}{affine} & \\
      \midrule
ballbeam & 2.141e-12 & 2.141e-12 & 1.746e-12\\
%batchProcessor-out1 & 1.187e-14 & 1.187e-14 & 1.048e-14\\
%batchProcessor-out2 & 3.187e-14 & 3.187e-14 & 2.911e-14\\
%batchProcessor-state1 & 7.902e-15 & 7.902e-15 & 7.132e-15\\
%batchProcessor-state2 & 8.521e-15 & 8.521e-15 & 7.840e-15\\
%batchProcessor-state3 & 8.405e-15 & 8.405e-15 & 7.689e-15\\
%batchProcessor-state4 & 7.008e-15 & 7.008e-15 & 6.217e-15\\
%batchReactor-out1 & 1.212e-15 & 1.212e-15 & 9.938e-16\\
%batchReactor-out2 & 3.141e-15 & 3.141e-15 & 2.692e-15\\
%batchReactor-state1 & 9.118e-16 & 9.118e-16 & 8.006e-16\\
%batchReactor-state2 & 8.451e-16 & 8.451e-16 & 7.455e-16\\
%batchReactor-state3 & 8.591e-16 & 8.591e-16 & 7.462e-16\\
%batchReactor-state4 & 7.693e-16 & 7.693e-16 & 7.019e-16\\
%bicycle-out1 & 6.943e-15 & 6.943e-15 & 5.450e-15\\
%bicycle-state1 & 5.927e-16 & 5.927e-16 & 4.778e-16\\
%bicycle-state2 & 4.815e-16 & 4.815e-16 & 3.995e-16\\
%bspline0 & 2.405e-16 & 2.405e-16 & 1.388e-16\\
bspline1 & 1.517e-15 & 1.601e-15 & 5.149e-16\\
bspline2 & 1.406e-15 & 1.448e-15 & 5.431e-16\\
bspline3 & 1.295e-16 & 1.295e-16 & 8.327e-17\\
%dcMotor-out1 & 7.851e-17 & 7.851e-17 & 6.419e-17\\
%dcMotor-state1 & 6.58e-16 & 6.58e-16 & 5.504e-16\\
%dcMotor-state2 & 1.172e-15 & 1.172e-15 & 9.978e-16\\
%dcMotor-state3 & 2.271e-17 & 2.271e-17 & 1.969e-17\\
doppler (m) & 9.766e-05 & 7.445e-04 & 3.111e-05\\
floudas1 & 1.052e-12 & 1.074e-12 & 5.755e-13\\
floudas26 & 7.292e-13 & 7.292e-13 & 7.740e-13\\
floudas33 & 3.109e-15 & 3.109e-15 & 6.199e-13\\
%floudas34 & 1.554e-15 & 1.554e-15 & 2.220e-15\\
%floudas46 & 2.798e-14 & 2.848e-14 & 1.554e-15\\
%floudas47 & 7.292e-13 & 7.292e-13 & 1.665e-14\\
himmilbeau (m) & 4.876e-04 & 4.876e-04 & 3.641e-04\\
invertedPendulum & 5.369e-14 & 5.369e-14 & 3.843e-14\\
kepler0 (m) & 2.948e-05 & 2.948e-05 & 1.758e-05\\
kepler1 (m) & 9.948e-05 & 9.948e-05 & 5.902e-05\\
kepler2 (m) & 3.732e-04 & 3.732e-04 & 1.433e-04\\
rigidBody1 (m) & 4.023e-05 & 4.023e-05 & 2.146e-05\\
rigidBody2 (m) & 6.438e-03 & 6.438e-03 & 9.871e-03\\
%verhulst & 8.343e-16 & 8.343e-16 & 3.235e-16\\
%predatorPrey & 3.395e-16 & 3.468e-16 & 1.836e-16\\
%carbonGas & 5.688e-08 & 5.666e-08 & 9.129e-09\\
traincar1-out1 & 5.406e-12 & 5.406e-12 & 4.601e-12\\
traincar1-state1 & 5.421e-15 & 5.421e-15 & 4.753e-15\\
traincar1-state2 & 8.862e-15 & 8.862e-15 & 8.099e-15\\
traincar1-state3 & 7.784e-15 & 7.784e-15 & 7.013e-15\\
%traincar2-out1 & 3.96e-12 & 3.96e-12 & 2.734e-12\\
%traincar2-state1 & 1.104e-14 & 1.104e-14 & 1.042e-14\\
%traincar2-state2 & 1.103e-14 & 1.103e-14 & 1.041e-14\\
%traincar2-state3 & 1.11e-14 & 1.11e-14 & 1.033e-14\\
%traincar2-state4 & 9.992e-15 & 9.992e-15 & 9.215e-15\\
%traincar2-state5 & 1.788e-14 & 1.788e-14 & 1.699e-14\\
%traincar3-out1 & 5.44e-11 & 5.44e-11 & 4.396e-11\\
%traincar3-state1 & 8.033e-15 & 8.033e-15 & 7.610e-15\\
%traincar3-state2 & 7.91e-15 & 7.91e-15 & 7.790e-15\\
%traincar3-state3 & 7.595e-15 & 7.595e-15 & 7.156e-15\\
%traincar3-state4 & 1.469e-14 & 1.469e-14 & 1.375e-14\\
%traincar3-state5 & 1.344e-14 & 1.344e-14 & 1.249e-14\\
%traincar3-state6 & 1.222e-14 & 1.222e-14 & 1.127e-14\\
%traincar3-state7 & 1.099e-14 & 1.099e-14 & 1.005e-14\\
%traincar4-out1 & 6.269e-10 & 6.269e-10 & 4.374e-10\\
%traincar4-state1 & 1.083e-14 & 1.083e-14 & 1.055e-14\\
%traincar4-state2 & 1.227e-14 & 1.227e-14 & 1.055e-14\\
%traincar4-state3 & 1.155e-14 & 1.155e-14 & 9.828e-15\\
%traincar4-state4 & 1.083e-14 & 1.083e-14 & 9.106e-15\\
%traincar4-state5 & 1.916e-14 & 1.916e-14 & 1.797e-14\\
%traincar4-state6 & 1.732e-14 & 1.732e-14 & 1.621e-14\\
%traincar4-state7 & 1.599e-14 & 1.599e-14 & 1.488e-14\\
%traincar4-state8 & 1.466e-14 & 1.466e-14 & 1.355e-14\\
%traincar4-state9 & 1.332e-14 & 1.332e-14 & 1.221e-14\\
turbine1 (m) & 1.356e-05 & 1.356e-05 & 3.192e-06\\
turbine2 (m) & 2.034e-05 & 2.034e-05 & 4.970e-06\\
turbine3 (m) & 9.038e-06 & 9.038e-06 & 1.671e-06\\
\bottomrule
    \end{tabular}
  \end{center}
  \caption{Roundoff errors verified by FloVer and FPTaylor.}\label{fig:error-results}
\end{table}
}

%%% Local Variables:
%%% mode: latex
%%% TeX-master: "paper"
%%% End:

%
\newcommand{\ctab}[1]{\multicolumn{1}{c}{#1}}
\begin{table}
  \footnotesize
  \begin{center}
    \renewcommand*{\arraystretch}{1.1}
    \setlength\tabcolsep{2pt}
    \begin{tabular}{lcr@{\hspace{0.75em}}|@{\hspace{0.75em}}rrrrr}
      \toprule
Benchmark & \# &  Daisy &\multicolumn{2}{c}{Coq} & HOL4 & CakeML & OCaml\\
 & ops &  & Interval & Affine &  &  &\\
      \midrule
ballBeam & 7 & 4.62 & 3.50 & 3.26 & 89.04 & <0.01 & 0.02\\
invertedPendulum & 7 & 3.62 & 3.59 & 3.27 & 112.61 & 0.01 & 0.02\\
bicycle & 13 & 4.31 & 4.01 & 4.08 & 156.76 & 0.01 & 0.04\\
doppler (m)& 17 & 4.86 & 5.28 & 12.21 & 610.67 & 0.05 & 0.02\\
dcMotor & 26 & 5.19 & 4.97 & 4.50 & 316.75 & 0.02 & 0.08\\
himmilbeau (m)& 26 & 3.52 & 4.11 & 4.40 & 65.48 & 0.02 & 0.03\\
bspline & 28 & 4.21 & 4.61 & 4.07 & 298.44 & 0.03 & 0.08\\
rigidbody (m)& 33 & 5.04 & 7.14 & 4.52 & 88.92 & 0.03 & 0.06\\
science & 35 & 5.64 & 11.69 & 567.36 & 1471.96 & 0.07 & 0.07\\
traincar1 & 36 & 4.85 & 10.87 & 9.84 & 932.93 & 0.07 & 0.11\\
batchProcessor & 56 & 6.46 & 8.49 & 7.43 & 997.77 & 0.06 & 0.16\\
batchReactor & 58 & 6.84 & 11.45 & 9.53 & 1117.48 & 0.07 & 0.17\\
turbine (m)& 82 & 5.99 & 18.69 & 24.90 & 4095.56 & 0.25 & 0.11\\
traincar2 & 89 & 7.90 & 29.79 & 28.58 & 3967.88 & 0.23 & 0.27\\
floudas & 99 & 7.76 & 13.99 & 12.76 & 565.68 & 0.14 & 0.27\\
kepler (m)& 158 & 4.89 & 21.56 & 22.70 & 3848.75 & 0.21 & 0.21\\
traincar3 & 168 & 9.14 & 68.53 & 68.14 & 9594.07 & 0.58 & 0.49\\
traincar4 & 269 & 10.6 & 116.94 & 115.38 & 17429.3 & 1.10 & 0.77\\
      \bottomrule
    \end{tabular}
  \end{center}
  \caption{Running times of Daisy and FloVer in seconds.}
  % The \coq{} and \hol{} columns show the runnning times for checking the certificate once inside the logic of the theorem prover.
  % The \daisy{} column is the average running time for parsing, analyzing and generating the certificate.
  % The binary columns show the running times for checking the static analysis using one of the extracted binaries a hundred times.
  % Running times are the `real' time of the UNIX \emph{time} command and have been rounded to the next full second.}
  \label{fig:eval}
\end{table}
To evaluate the performance of FloVer, we have extended the static analyzer
Daisy to generate certificates of its analysis.
As Daisy already computes all the information that needs to be encoded in a
certificate, implementing the certificate generation was similar to
implementing a pretty-printer for analysis results (we have switched off
a few optimizations, which however do not affect the error bounds significantly).
Using the certificate generation, we have evaluated Daisy and FloVer on examples
taken from the Rosa~\cite{Rosa2015} and real2float~\cite{Magron2015} projects.
Each benchmark consists of one or more separate functions. Daisy analyzes all
functions of one benchmark together and produces one certificate containing a call to the
certificate checker for each separate function.

We compare error bounds verified by FloVer with those
verified by FPTaylor, as FPTaylor generally computes the most accurate
bounds~\cite{Solovyev2015,Darulova2016}. Furthermore,
Rosa~\cite{Rosa2015}, Fluctuat~\cite{Goubault2013} and Gappa~\cite{Daumas2010}
use the same technique to compute roundoff errors as Daisy and FloVer.
We also compare FloVer's certificate checking times with FPTaylor's, as the tool
also provides a proof certificate.%and not a proof script.
We note that FPTaylor can compute less precise error bounds with shorter running times,
here we opt for the off-the-shelve solution without additional parameters.
% Our efficiency evaluation presents only FloVer's running times. We give the
% running times for FPTaylor's certificate checking in the
% \hyperref[sec:appendix]{appendix} and note that they are in the same orders of
% magnitude as our Coq in-logic evaluation.
% Our fixed-point checking has similar running times as our floating-point
% checking and we give running times for checking fixed-point programs in the
% \hyperref[sec:appendix]{appendix}.
% Our efficiency evaluation compares only FloVer's running times, as FPTaylor
% has certificate checking times similar to FloVer's while requiring either a
% two hour startup time, or external checkpointing.

\paragraph{Accuracy}
%To show that FloVer is able to prove reasonable error bounds,
\autoref{fig:error-results} gives a subset of the roundoff errors certified by
FloVer as well as roundoff errors computed by FPTaylor for
comparison (we give the full table in the appendix (\autoref{sec:appendix}).%our technical report~\cite{becker2017verified}).
PRECiSa and Gappa compute similar results; we provide them here for two benchmarks for reference.
For the ballBeam benchmark, Precisa and Gappa show an error of 1.085e-07 and
1.240e-12 resp., and for the invertedPendulum benchmark, the errors are
3.531e-12 and 3.217e-14.
%\heiko{Is this the correct place to add the Precisa/Gappa numbers? Should we have more?}
The focus of FloVer is not to compute the most precise bounds possible, but rather to
develop the necessary infrastructure for future extensions.
Nevertheless, the roundoff errors verified by it are usually close to those
proven by FPTaylor.
% shows the roundoff errors for all benchmarks that
% have been certified by FloVer during our evaluation.
% The focus of FloVer is not to produce tight bounds, but rather to verify bounds
% computed by other tools, thus we give only the bounds computed by FPTaylor as a
% reference, since FPTaylor is in general able to compute very tight bounds.
Benchmarks marked with `(m)' are in mixed-precision,
otherwise the roundoff errors are evaluated under uniform double (64 bit)
floating-point precision (FPTaylor does not support fixed-point precision).

\paragraph{Efficiency}
In~\autoref{fig:eval}, we compare running times of in-logic evaluation of
FloVer in Coq and HOL4, the \emph{verified} binary extracted with the CakeML
toolchain and the \emph{unverified} binary extracted from \coq{}. For
our experiments we used a machine with a four core Intel i3 processor with
3.3GHz, 8 GB of RAM, running Debian 9. For the in-logic evaluation in Coq we
show range analysis in interval and affine arithmetic, for all other runs we
use interval arithmetic. As for the accuracy evaluation, benchmarks marked with
`(m)' are in mixed-precision, double precision otherwise.

In~\autoref{fig:eval}, `OCaml' refers to the Coq binary compiled with the
OCaml native compiler. The `\# ops' column gives the number of arithmetic
operations in the whole benchmark (summed for all functions) and gives an
intuition about the complexity of the benchmark.
For all columns, the running times are the end-to-end times measured by the
UNIX \emph{time} command in seconds. This time includes parsing and generating the
certificate for Daisy, checking the proof that FloVer succeeds for Coq and
HOL4 in-logic, and running FloVer in the binaries.
The running times for Daisy, \coq{} and \hol{} are the average running times
for a single run over three runs. For the binaries we report the average
running time of a single run from 300 executions (due to the small runtime).

% No direct comparison
We give the running times for FPTaylor's certificate checking in the appendix (\autoref{sec:appendix})%our technical
%report~\cite{becker2017verified}
and note that they are larger, but of the same order of magnitude
as our Coq in-logic evaluation. Note that FPTaylor's checker
requires either a two hour starting time or external checkpointing.
FloVer's certificate checking time for fixed-point arithmetic is similar to floating-point
checking; we give the detailed running times in the appendix (\autoref{sec:appendix}).%our technical report~\cite{becker2017verified}.

% We do not directly compare FloVer to Gappa or PRECiSA, but give the running
% times for FPTaylor in the \hyperref[sec:appendix]{appendix}.
% If we ignore that FPTaylor's checker requires either a two hour starting time or
% external checkpointing, its running time is in the same orders of magnitude as FloVer's in Coq.
% Furthermore FloVer's focus is first on ease of use and then on performance.
% % FPTaylor needs up to 2h startup time
% % generate scripts, and all only support in-logic
% We do not compare FloVer's running times against other tools,
% as Gappa and PRECiSA generate proof scripts and require user-interaction, and FPTaylor
% requires a start-up time of up to 2h. Furthermore all other tools only
% support in-logic verification.

% % Since the execution times of Daisy, \coq{} and \hol{} are substantial,
% % we report them for a single run. For the binaries however, the execution times
% % are much smaller, so that we report the times for checking the certificates
% % hundred times.

%\heiko{Below ok as rewrite?}
The evaluation of FloVer's Coq checker is faster than the evaluation of the HOL4
checker. This is probably because we benefit from Coq's \verb!vm_compute! tactic
in the Coq evaluation. The tactic translates terms to OCaml and evaluates
them using a virtual machine. A Coq term is reconstructed from the result.
HOL4's \verb!EVAL_TAC! instead uses a simple call-by-value evaluation strategy.
% This is probably due to the fact that evaluation in \hol{} still passes
% through the kernel in the form of inference rule applications. In \coq{} the
% functions that are evaluated need not be checked by the kernel when being run,
% apart from type-checking the arguments.
We further observe that the evaluation using affine arithmetic sometimes is as
fast as the one using intervals. We suspect that the reason for this is that the
affine arithmetic checker must memorize polynomials for sub-expressions and thus
does not recompute them. The interval validator, however, currently does not
memorize sub-expressions, but only let-bound variables.
\section{Related Work}
% We first review the most related work in accuracy analysis of floating-point
% arithmetic computations and then put our work into a broader context.

\paragraph{Sound Accuracy Analysis}
The tools FPTaylor~\cite{Solovyev2015}, Gappa~\cite{Daumas2010},
PRECiSa~\cite{moscatostatic}, real2float~\cite{Magron2015} and
VCFloat~\cite{Ramananandro2016} are most closely related to our work as they
formally verify floating-point roundoff errors.
Each tool handles mixed-precision floating-point arithmetic, but
other features differ slightly between tools.
FloVer is the only tool with the combination of support for both \coq{} and
\hol{}, floating-point as well as fixed-point arithmetic and two abstract
domains, interval and affine arithmetic. %(full support for AA is our priority work in progress).
FloVer is fully automated and FloVer and FPTaylor are the only tools that
generate certificates using in-logic decision procedures %and do not require proof scripts.
While FPTaylor and PRECiSa handle transcendental functions (which FloVer does not), both tools do
not handle fixed-point arithmetic.
Gappa has some support for fixed-points, but FloVer is the only tool with
formalized affine arithmetic.
Finally, FloVer is the first tool to provide \emph{efficient} certificate
checking with a verified binary.
%
%
% {\small
% \begin{table*}[tb]
%   \begin{center}
%     \renewcommand*{\arraystretch}{1.2}
%     \begin{tabular}{@{}clllllllll@{}}
%       \toprule
%       Tool & FloVer & FPTaylor~\cite{Solovyev2015}
%       & Gappa~\cite{Daumas2010}
%       & Precisa~\cite{moscatostatic}
%       & VCFloat\cite{Ramananandro2016}\\
%       \midrule
%       Supported Prover & Coq, HOL4 & HOL-Light & Coq & PVS & Coq\\
%       Automation & Full & Full & Partial (user hints possible) & Full &Partial (automation tactics)\\
%       IEEE Connection & Yes & Yes & Yes & Yes & Yes\\
%       Control Flow & let's & let's & let's & let's$+$ & let's$+$\\
%       Verification Method & IA, AA & Taylor Appr. & IA & Denot. Semantics & Syntactic Analysis\\
%       Transcendental Functions & No & Yes & No & Yes & No\\
%       Mixed Precision & Yes & Yes & Yes & Yes & Yes\\
%       Verified Binary & Yes & No & No & No & No\\
%       Input Language & Coq, HOL4 or custom & Custom & Custom & PVS files & CompCert C-light\\
%       \bottomrule
%     \end{tabular}
%   \end{center}
%   \caption{Comparison of supported features of related tools}
%   \label{fig:rel_overview}
% \end{table*}
% }
%
Fluctuat~\cite{Goubault2011}, Gappa++~\cite{linderman2010towards} and
Rosa~\cite{Darulova2016} statically bound finite-precision roundoff errors
using affine arithmetic~\cite{Figueiredo2004}, but
do not provide formal guarantees.
%  All of these tools support affine
% arithmetic~\cite{Figueiredo2004}, as we do in FloVer.

FloVer currently does not handle conditionals and loops. These are---to some
extent---supported by Fluctuat~\cite{Goubault2013} and
Rosa~\cite{Darulova2016}, however not formally verified.
PRECiSa~\cite{moscatostatic} provides an initial formalization of these
approaches, but scalability is unclear~\cite{Rosa2015,Darulova2016}.
%
% Roundoff error analysis for programs with loops is challenging, because roundoff
% errors in general grow with each loop iteration and thus a nontrivial
% fixpoint does not exist in general. Fluctuat and Precisa provide widening operators to
% support loops, however these compute non-trivial bounds only for very special
% cases where roundoff errors decrease with each loop iteration. A (not formally
% verified) technique for more general loops has been presented in
% Rosa~\cite{Darulova2016}, however it also makes additional assumptions which
% only apply to certain types of loops. We focus in this work on a general fully
% automated technique and leave a proper treatment of loops and conditionals to
% future work. We note, however, that with loop unrolling, FloVer can already be
% applied to programs with loops today.
%
FloVer furthermore focuses, like most tools, on certifying absolute error bounds.
Bounding relative errors is challenging due to the increased complexity as well
as due to the issue that often the error is not even well-defined due to an
inherent division by zero~\cite{Izycheva2017}.
Gappa does provide verified relative error support by optimizing a constraint
based on~\autoref{eqn:floats}. This approach has been shown to not provide
tight bounds once input ranges and expressions become larger~\cite{Izycheva2017}.
Finally, note that input ranges are also necessary for
computing \emph{concrete} relative error bounds.

% FloVer checks certificates for absolute errors. An automated and general estimation
% of relative errors ($\lvert f(x) - \tilde{f}(\tilde{x})
% \rvert / \lvert f(x)\rvert$), though it may be more desirable, presents a
% significant challenge today~\cite{Izycheva2017}. For instance, when the range of
% the expression in question (i.e. the range of $f(x)$) includes zero,
% relative errors are not well defined. Unfortunately, this is often the case
% in practice.
% \heiko{This could potentially be merged with the related work discussion}

\paragraph{Sound Verification of Floating-point Computations}
Absence of runtime errors in floating-point computations can be shown with
abstract interpretation, where different abstract domains have been developed
for this purpose~\cite{Blanchet2003,Chen2008,Jeannet2009}, which are sound
w.r.t. floating-point arithmetic. Jourdan et al.~\cite{Jourdan2015} have also
formalized some of these abstract domains in Coq. Note, however, that these
domains do not quantify the difference between a real-valued and the
finite-precision semantics and can only show the absence of runtime errors.

Moscato et al.~\cite{MoscatoMS15} have built a formalization and implementation
of AA for computation of real-valued ranges in PVS.
This development does not handle division, which we do.
Immler~\cite{immler2015verified} has formalized AA in Isabelle/HOL;
our own formalization shares a similar structure.
% but is used for bounding real-valued
% ranges instead of computing the join of two affine polynomials.
% \heiko{Ok? Hyperplane intersection should be the join for affine polynomials, right? }
% Our focus was not to compute the tightest possible bounds but on modularity and
% reusability of the framework. Nevertheless, we may be able to use some of their
% ideas when extending our framework to affine arithmetic.

Coq has also been used to prove entire programs correct w.r.t. numerical
uncertainties such as roundoff errors~\cite{Boldo2013}.
However, in these efforts much of the work is still manual. Our current
development can be seen as complementary as it could potentially provide
automation for the verification of roundoff error bounds.
The CompCert compiler also supports floating-point computations~\cite{Boldo2015a},
but only shows semantics preservation and not roundoff error bounds.
Harrison~\cite{harrison1997} has formally verified a floating point implementation
of the exponential function inside HOL-Light. The analysis is detailed and specific to this
particular function. In contrast, our work aims to provide a fully automated verified
analysis for arbitrary real-valued expressions, but at a higher level of abstraction.

% In the context of automated verification within SMT-solvers,
% R\"ummer and Wahl~\cite{Rummer2010} have proposed an SMT-lib theory for floating-point arithmetic.
% Decision procedures for this theory mostly rely on a bit-precise encoding and in
% general need approximation techniques to deal with the inherent prohibitive
% complexity, see for instance~\cite{Haller2012}. Again, these techniques do not estimate roundoff errors wrt.
% a real-valued semantics, in part because this require a combination of theories
% which is an open problem at present.

\paragraph{Real Arithmetic and Finite-precision Formalizations}\hspace{1pt}\\
Formalizations of floating-point arithmetic exist in HOL-Light~\cite{Jacobsen2015},
in Coq in the Flocq library~\cite{Boldo2011} as well as in Isabelle~\cite{IEEE_FP_AFP} and \hol{}~\cite{IEEE_FP_HOL4}.
% All of these are able to closely model the hardware specification of floating
% points, for instance, they provide a representation which can handle different
% rounding modes and expresses floating point values using a mantissa and an exponent.
We found using these formalizations in \coq{} and \hol{} more complex than was
necessary for reasoning inside FloVer, thus we use them only to show a connection to IEEE754.
Fixed-point arithmetic has been formalized in HOL4~\cite{Akbarpour2005}, focusing
on its hardware implementation,
whereas our focus is on relating their execution to real-valued semantics.
%our purpose so that we relied on the simpler abstraction from~\autoref{eqn:floats}, apart from using the IEEE754
%\todo[inline]{Heiko: Does the above make clear what we did?}%Heiko: Raphael suggested rephrasing, as we now use the IEEE libs}

% \eva{CUT? (is this really relevant?) To the best of our knowledge, there exist two frameworks which extend \coq{}
% with formalisations of real arithmetic: the Coquelicot library from
% Boldo et al.~\cite{Boldo2015} and the C-CoRN library~\cite{Robbert2011}.
% Since we made the sound simplification of using rationals for our computations, it was
% not necessary to use either \coq{} formalisation of the real numbers in our development.
% Boldo et al.~\cite{boldo2014} provide a broader comparison of theorem
% provers in the context of real numbers, e.g. their implementation, inclusion of
% certain operations and support for proof automation.}

%Our intention is to provide a comparison between \coq{} and \hol{} for the special case of
%verification of numerical error bounds.
%

%%% Local Variables:
%%% mode: latex
%%% TeX-master: "paper"
%%% End:

%
% !TEX root = paper.tex

% \item Conclusion:
%   We have presented a new FP analysis framework, Daisy, which can compute and fully automatically verify error bounds of floating point programs.
%    The analysis and the validation are both modular to allow for easy extensions
%  \item Future Work: Fully certified analysis pipeline using CompCert for Coq and cakeML for HOL4
\section{Conclusion}
We have presented our modular, reusable and easily extendable approach to
certificate checking for error bound analysis in FloVer. Our checker is
fully-automated and requires neither user interaction, nor expert knowledge.
All of the theorems about FloVer have been proven in both \coq{} and \hol{}.
We are the first to extract a verified binary for checking finite-precision
roundoff errors using the CakeML toolchain
% extractor to extract a verified binary from
% \hol{} code
and have shown that we achieve significant performance improvements
when using the binary.
%\todo[inline]{Heiko: Added 1 sentence about what we have proven, as suggested by Anthony Fox.}
% \todo[inline]{Eva: now it says that noone has used the CakeML extractor before - on anything.
% Is this intended?}
% \todo[inline]{Eva: perhaps we can say here that FloVer could also be used to
% verify results produced from potentially unsound, but more efficient analyses.
% Such analyzes could be interesting for instance for optimization techniques,
% which need to evaluate roundoff errors frequently.}

% \paragraph*{\textbf{\ackname}}
% The authors would like to thank Jacques-Henri Jourdan for helping to solve practical problems with our Coq development and the many insightful discussions.
%%% Local Variables:
%%% mode: latex
%%% TeX-master: "paper"
%%% End:

%equalize columns

%% Bibliography
\bibliography{references}

\clearpage
\newpage
\appendix
\section{Appendix}
\label{sec:appendix}

%\subsection{FloVer and FPTaylor Roundoff Errors}
\renewcommand*{\arraystretch}{1.1}

\tablefirsthead{
  \toprule
  Benchmark & \multicolumn{2}{c}{FloVer} & FPTaylor\\
   & \multicolumn{1}{c}{interval} & \multicolumn{1}{c}{affine} & \\
  \midrule}
\tablehead{%
%\multicolumn{2}{c}%
%{{Continued from previous column}} \\
  \toprule
  Benchmark & \multicolumn{2}{c}{FloVer} & FPTaylor\\
   & \multicolumn{1}{c}{interval} & \multicolumn{1}{c}{affine} & \\
  \midrule}
\tabletail{%
\midrule}% \multicolumn{2}{r}{{Continued on next column}} \\ \midrule}
\tablelasttail{\\ \bottomrule}
\tablecaption{Full table with all the roundoff errors verified by FloVer and
  computed by FPTaylor from our evaluation in \autoref{sec:experiments}.}
{\small
\begin{supertabular}{lrrr}
  \shrinkheight{40pt}
ballbeam & 2.141e-12 & 2.141e-12 & 1.746e-12\\
batchProcessor-out1 & 1.187e-14 & 1.187e-14 & 1.048e-14\\
batchProcessor-out2 & 3.187e-14 & 3.187e-14 & 2.911e-14\\
batchProcessor-state1 & 7.902e-15 & 7.902e-15 & 7.132e-15\\
batchProcessor-state2 & 8.521e-15 & 8.521e-15 & 7.840e-15\\
batchProcessor-state3 & 8.405e-15 & 8.405e-15 & 7.689e-15\\
batchProcessor-state4 & 7.008e-15 & 7.008e-15 & 6.217e-15\\
batchReactor-out1 & 1.212e-15 & 1.212e-15 & 9.938e-16\\
batchReactor-out2 & 3.141e-15 & 3.141e-15 & 2.692e-15\\
batchReactor-state1 & 9.118e-16 & 9.118e-16 & 8.006e-16\\
batchReactor-state2 & 8.451e-16 & 8.451e-16 & 7.455e-16\\
batchReactor-state3 & 8.591e-16 & 8.591e-16 & 7.462e-16\\
batchReactor-state4 & 7.693e-16 & 7.693e-16 & 7.019e-16\\
bicycle-out1 & 6.943e-15 & 6.943e-15 & 5.450e-15\\
bicycle-state1 & 5.927e-16 & 5.927e-16 & 4.778e-16\\
bicycle-state2 & 4.815e-16 & 4.815e-16 & 3.995e-16\\
bspline0 & 2.405e-16 & 2.405e-16 & 1.388e-16\\
bspline1 & 1.517e-15 & 1.601e-15 & 5.149e-16\\
bspline2 & 1.406e-15 & 1.448e-15 & 5.431e-16\\
bspline3 & 1.295e-16 & 1.295e-16 & 8.327e-17\\
dcMotor-out1 & 7.851e-17 & 7.851e-17 & 6.419e-17\\
dcMotor-state1 & 6.58e-16 & 6.58e-16 & 5.504e-16\\
dcMotor-state2 & 1.172e-15 & 1.172e-15 & 9.978e-16\\
dcMotor-state3 & 2.271e-17 & 2.271e-17 & 1.969e-17\\
doppler (m) & 9.766e-05 & 7.445e-04 & 3.111e-05\\
floudas1 & 7.292e-13 & 7.292e-13 & 5.755e-13\\
floudas26 & 1.052e-12 & 1.074e-12 & 7.740e-13\\
floudas33 & 7.292e-13 & 7.292e-13 & 6.199e-13\\
floudas34 & 3.109e-15 & 3.109e-15 & 2.220e-15\\
floudas46 & 1.554e-15 & 1.554e-15 & 1.554e-15\\
floudas47 & 2.798e-14 & 2.848e-14 & 1.665e-14\\
himmilbeau (m) & 4.876e-04 & 4.876e-04 & 3.641e-04\\
invertedPendulum & 5.369e-14 & 5.369e-14 & 3.843e-14\\
kepler0 (m) & 2.948e-05 & 2.948e-05 & 1.758e-05\\
kepler1 (m) & 9.948e-05 & 9.948e-05 & 5.902e-05\\
kepler2 (m) & 3.732e-04 & 3.732e-04 & 1.433e-04\\
rigidBody1 (m) & 4.023e-05 & 4.023e-05 & 2.146e-05\\
rigidBody2 (m) & 1.288e-02 & 1.288e-02 & 9.871e-03\\
verhulst & 8.343e-16 & 8.343e-16 & 3.235e-16\\
predatorPrey & 3.395e-16 & 3.468e-16 & 1.836e-16\\
carbonGas & 5.688e-08 & 5.666e-08 & 9.129e-09\\
traincar1-out1 & 5.406e-12 & 5.406e-12 & 4.601e-12\\
traincar1-state1 & 5.421e-15 & 5.421e-15 & 4.753e-15\\
traincar1-state2 & 8.862e-15 & 8.862e-15 & 8.099e-15\\
traincar1-state3 & 7.784e-15 & 7.784e-15 & 7.013e-15\\
traincar2-out1 & 3.96e-12 & 3.96e-12 & 2.734e-12\\
traincar2-state1 & 1.104e-14 & 1.104e-14 & 1.042e-14\\
traincar2-state2 & 1.103e-14 & 1.103e-14 & 1.041e-14\\
traincar2-state3 & 1.11e-14 & 1.11e-14 & 1.033e-14\\
traincar2-state4 & 9.992e-15 & 9.992e-15 & 9.215e-15\\
traincar2-state5 & 1.788e-14 & 1.788e-14 & 1.699e-14\\
traincar3-out1 & 5.44e-11 & 5.44e-11 & 4.396e-11\\
traincar3-state1 & 8.033e-15 & 8.033e-15 & 7.610e-15\\
traincar3-state2 & 7.91e-15 & 7.91e-15 & 7.790e-15\\
traincar3-state3 & 7.595e-15 & 7.595e-15 & 7.156e-15\\
traincar3-state4 & 1.469e-14 & 1.469e-14 & 1.375e-14\\
traincar3-state5 & 1.344e-14 & 1.344e-14 & 1.249e-14\\
traincar3-state6 & 1.222e-14 & 1.222e-14 & 1.127e-14\\
traincar3-state7 & 1.099e-14 & 1.099e-14 & 1.005e-14\\
traincar4-out1 & 6.269e-10 & 6.269e-10 & 4.374e-10\\
traincar4-state1 & 1.083e-14 & 1.083e-14 & 1.055e-14\\
traincar4-state2 & 1.227e-14 & 1.227e-14 & 1.055e-14\\
traincar4-state3 & 1.155e-14 & 1.155e-14 & 9.828e-15\\
traincar4-state4 & 1.083e-14 & 1.083e-14 & 9.106e-15\\
traincar4-state5 & 1.916e-14 & 1.916e-14 & 1.797e-14\\
traincar4-state6 & 1.732e-14 & 1.732e-14 & 1.621e-14\\
traincar4-state7 & 1.599e-14 & 1.599e-14 & 1.488e-14\\
traincar4-state8 & 1.466e-14 & 1.466e-14 & 1.355e-14\\
traincar4-state9 & 1.332e-14 & 1.332e-14 & 1.221e-14\\
turbine1 (m) & 1.356e-05 & 1.356e-05 & 3.192e-06\\
turbine2 (m) & 2.034e-05 & 2.034e-05 & 4.970e-06\\
turbine3 (m) & 9.038e-06 & 9.038e-06 & 1.671e-06\\
\end{supertabular}
}
\clearpage
\newpage
\tablefirsthead{
  \toprule
  Benchmark & FPTaylor & Coq In-logic\\
  \midrule}
\tablehead{%
%\multicolumn{2}{c}%
%{{Continued from previous column}} \\
  \toprule
  Benchmark & FPTaylor & Coq In-logic\\
  \midrule}
\tabletail{%
\midrule}% \multicolumn{2}{r}{{Continued on next column}} \\ \midrule}
\tablelasttail{\\ \bottomrule}
\tablecaption{This table gives FPTaylor's running times measured by the ocaml `time' command
in seconds.
The numbers are measured \emph{after} checkpointing a HOL-Light state with the
FPTaylor solver loaded and thus include only certificate checking times. Each
number is the average running time of a single run measured over three runs.
Since FPTaylor produces separate certificates for each function in the input
file, we report them separately.
`Error' refers to FPTaylor's checker failing with the exception \lstinline{Invalid_argument "index out of bounds"}.
For comparison, we give the Coq in-logic checking times for all functions in the
benchmark file, as they are closest to those of FPTaylor.}
{\small
\begin{supertabular}{lrr}
  \shrinkheight{70pt}
ballbeam & 12.37 & 3.50\\
\midrule
batchProcessor-out1 & 19.54 & \multirow{6}{*}{7.28}\\
batchProcessor-out2 & 19.51 &\\
batchProcessor-state1 & 27.33 &\\
batchProcessor-state2 & 37.84 &\\
batchProcessor-state3 & 37.33 &\\
batchProcessor-state4 & 37.74 &\\
\midrule
batchReactor-out1 & 19.11 & \multirow{6}{*}{12.03}\\
batchReactor-out2 & 18.82 &\\
batchReactor-state1 & 37.94 &\\
batchReactor-state2 & 38.69 &\\
batchReactor-state3 & 38.73 &\\
batchReactor-state4 & 38.33 &\\
\midrule
bicycle-out1 & 6.37 & \multirow{3}{*}{4.03}\\
bicycle-state1 & 11 &\\
bicycle-state2 & 10.92 &\\
\midrule
bspline0 & 10.77 & \multirow{4}{*}{4.46}\\
bspline1 & 15.67 &\\
bspline2 & 19.26 &\\
bspline3 & 6.38 &\\
\midrule
dcMotor-out1 & 11.56 & \multirow{4}{*}{5.13}\\
dcMotor-state1 & 17.82 &\\
dcMotor-state2 & 17.57 &\\
dcMotor-state3 & 17.44 &\\
\midrule
doppler (m) & 231.37 & 6.10\\
\midrule
floudas1 & 63.2 & \multirow{6}{*}{13.16}\\
floudas26 & Error &\\
floudas33 & 71.15 &\\
floudas34 & 4.85 &\\
floudas46 & 2.68 &\\
floudas47 & 7.9 &\\
himmilbeau (m) & 49.95 & 4.92\\
\midrule
invertedPendulum & 15.91 & 3.64\\
\midrule
kepler0 (m) & 70.42 & \multirow{3}{*}{20.51}\\
kepler1 (m) & 147.18 &\\
kepler2 (m) & 412.96 &\\
\midrule
rigidBody1 (m) & 14.18 & \multirow{2}{*}{4.90}\\
rigidBody2 (m) & 47.33 &\\
\midrule
verhulst & 12.21 & \multirow{3}{*}{9.70}\\
predatorPrey & 28.99 &\\
carbonGas & 28.21 &\\
\midrule
traincar1-out1 & 15.31 & \multirow{4}{*}{10.25}\\
traincar1-state1 & 46.64 &\\
traincar1-state2 & 45.59 &\\
traincar1-state3 & 47.57 &\\
\midrule
traincar2-out1 & 31.94 & \multirow{6}{*}{28.92}\\
traincar2-state1 & 81.43 &\\
traincar2-state2 & 84.77 &\\
traincar2-state3 & 86.59 &\\
traincar2-state4 & 81.17 &\\
traincar2-state5 & 74.61 &\\
\midrule
traincar3-out1 & Error & \multirow{8}{*}{66.90}\\
traincar3-state1 & Error &\\
traincar3-state2 & Error &\\
traincar3-state3 & Error &\\
traincar3-state4 & Error &\\
traincar3-state5 & Error &\\
traincar3-state6 & Error &\\
traincar3-state7 & Error &\\
\midrule
traincar4-out1 & Error & \multirow{10}{*}{112.95}\\
traincar4-state1 & Error &\\
traincar4-state2 & Error &\\
traincar4-state3 & Error &\\
traincar4-state4 & Error &\\
traincar4-state5 & Error &\\
traincar4-state6 & Error &\\
traincar4-state7 & Error &\\
traincar4-state8 & Error &\\
traincar4-state9 & Error &\\
\midrule
turbine1 (m) & 80.96 & \multirow{3}{*}{18.38}\\
turbine2 (m) & 56.41 &\\
turbine3 (m) & 82.86 &\\
\end{supertabular}
}

\clearpage
\begin{table*}[!htp]
  \centering
  \caption{This table shows our evaluation of FloVer on fixed-point arithmetic.
For this we used Daisy's fixed-point analysis, to generate a certificate for a
word length of 32 bits. `\# ops' is the numbers of arithmetic operations in the
file which we use as our complexity measure. As for our evaluation in
\autoref{sec:experiments} all measurements are elapsed time from the UNIX
\emph{time} command, in seconds.}
  \begin{tabular}{lrrrrrr}
    \toprule
    Benchmark & \# ops & Daisy & Coq (IA)& HOL4 & CakeML & OCaml\\
    \midrule
    ballBeam & 7 & 3.25 & 3.07 & 51.03 & <0.01 & <0.01\\
    invertedPendulum & 7 & 3.17 & 3.03 & 57.55 & <0.01 & <0.01\\
    bicycle & 13 & 3.48 & 3.27 & 73.98 & <0.01 & <0.01\\
    doppler & 17 & 3.14 & 3.21 & 73.14 & <0.01 & <0.01\\
    dcMotor & 26 & 3.81 & 3.86 & 94.69 & <0.01 & 0.01\\
    himmilbeau & 26 & 2.99 & 3.11 & 50.30 & <0.01 & <0.01\\
    bspline & 28 & 3.59 & 3.50 & 74.84 & <0.01 & 0.01\\
    rigidbody & 33 & 3.32 & 3.36 & 58.80 & <0.01 & 0.01\\
    science & 35 & 3.54 & 4.11 & 180.2 & 0.01 & <0.01\\
    traincar1 & 36 & 4.11 & 6.48 & 252.19 & 0.01 & 0.02\\
    batchProcessor & 56 & 4.55 & 4.67 & 153.17 & 0.01 & 0.02\\
    batchReactor & 58 & 4.61 & 5.39 & 189.14 & 0.01 & 0.02\\
    turbine & 82 & 3.88 & 5.20 & 325.77 & 0.03 & 0.01\\
    traincar2 & 89 & 5.36 & 14.37 & 622.37 & 0.03 & 0.04\\
    floudas & 99 & 5.54 & 6.27 & 125.58 & 0.01 & 0.04\\
    kepler & 158 & 3.92 & 6.24 & 216.79 & 0.01 & 0.01\\
    traincar3 & 168 & 6.47 & 31.01 & 1244.96 & 0.07 & 0.10\\
    traincar4 & 269 & 7.72 & 45.08 & 1479.48 & 0.09 & 0.18\\
    \bottomrule
  \end{tabular}
\end{table*}
%%% Local Variables:
%%% mode: latex
%%% TeX-master: "paper"
%%% End:

% % use section* for acknowledgment
% \section*{Acknowledgment}
\end{document}